\documentclass[sigconf]{acmart}

\def\BibTeX{{\rm B\kern-.05em{\sc i\kern-.025em b}\kern-.08emT\kern-.1667em\lower.7ex\hbox{E}\kern-.125emX}}

\acmConference[ICSE 2024]{46th International Conference on Software Engineering}{April 2024}{Lisbon, Portugal}

\PassOptionsToPackage{utf8}{inputenc}
\usepackage{inputenc}

\providecommand{\mLyX}{L\kern-.1667em\lower.25em\hbox{Y}\kern-.125emX\@}

\usepackage{csquotes}
 
\PassOptionsToPackage{fleqn}{amsmath} 
 \usepackage{amsmath}
 
\usepackage{xspace} 

\PassOptionsToPackage{smaller}{acronym} 
\usepackage[nolist]{acronym} 

\PassOptionsToPackage{pdftex}{graphicx}
\usepackage{graphicx} 

\usepackage{tabularx} 
\newcommand{\myfloatalign}{\centering} 
\usepackage{caption}
\usepackage{subcaption}
\usepackage{xcolor}

\usepackage{listings} 
\makeatletter
\@ifpackageloaded{babel}
{
\addto\extrasamerican{

}
}{\relax}
\makeatother

\lstdefinelanguage{JavaScript}{
  keywords={constructor, typeof, new, true, false, catch, function, return, null, catch, switch, var, if, in, while, do, else, case, break, let},
  keywordstyle=\color{RoyalBlue},
  ndkeywords={class, export, boolean, throw, implements, import, this},
  ndkeywordstyle=\color{RoyalBlue},
  identifierstyle=\color{black},
  sensitive=false,
  commentstyle=\color{ForestGreen}\ttfamily,
  comment=[l]{//},
  morecomment=[s]{/*}{*/},
  stringstyle=\color{red}\ttfamily,
  morestring=[b]',
  morestring=[b]"
}

\usepackage[capitalise]{cleveref}

\usepackage{float}

\usepackage{scratch3}
\setscratch{scale=.5}

\usepackage[ruled,vlined]{algorithm2e}
\SetKwInOut{Input}{input}
\SetKwInOut{Output}{output}
\SetKwProg{Fn}{function}{:}{end}
\SetFuncSty{textsf}

\newcommand{\inlineFigure}[2]{$\vcenter{\hbox{\includegraphics[scale=#1]{#2}}}$}

\newcolumntype{L}[1]{>{\raggedright\let\newline\\\arraybackslash\hspace{0pt}}m{#1}}
\newcolumntype{C}[1]{>{\centering\let\newline\\\arraybackslash\hspace{0pt}}m{#1}}
\newcolumntype{R}[1]{>{\raggedleft\let\newline\\\arraybackslash\hspace{0pt}}m{#1}}

\newcommand{\textmessage}[1]{\textls[20]{\textit{{#1}}}}
\newcommand{\textmessagetable}[1]{\textls[20]{\textit{{#1}}}}
\newcommand{\textcode}[1]{\textls[20]{\textsf{{#1}}}}

\newcommand\definetool[2]{\newcommand{#1}{{\textsc{#2}}\xspace}}
\definetool{\Scratch}{Scratch}
\definetool{\toolname}{NuzzleBug}
\definetool{\whyline}{Whyline}
\definetool{\alice}{Alice}
\definetool{\snap}{Snap!}
\definetool{\pencilcode}{PencilCode}
\definetool{\makecode}{MakeCode}
\definetool{\blockly}{Blockly}
\definetool{\java}{Java}
\definetool{\JavaScript}{JavaScript}
\definetool{\React}{React}
\definetool{\Whisker}{Whisker}
\definetool{\LitterBox}{LitterBox}

\newcommand{\WhyDid}{\textmessage{``Why did ...?''}\xspace}
\newcommand{\WhyDidnt}{\textmessage{``Why didn't ...?''}\xspace}
\newcommand{\WhenDid}{\textmessage{``When did ...?''}\xspace}

\definecolor{lightgray}{rgb}{.8,.8,.8}
\definecolor{darkgray}{rgb}{.4,.4,.4}

\usepackage{fdsymbol}

\newcommand{\summary}[2]{%
	\vspace{-0.3cm}%
	\begin{center}%
		\colorbox{gray!20}{%
			\parbox{\linewidth}{%
				\textbf{\textsf{Summary (\textit{#1})}:}~%
				#2%
			}%
		}%
	\end{center}%
}

\copyrightyear{2024}
\acmYear{2024}
\setcopyright{acmlicensed}\acmConference[ICSE '24]{2024 IEEE/ACM 46th International Conference on Software Engineering}{April 14--20, 2024}{Lisbon, Portugal}
\acmBooktitle{2024 IEEE/ACM 46th International Conference on Software Engineering (ICSE '24), April 14--20, 2024, Lisbon, Portugal}
\acmPrice{15.00}
\acmDOI{10.1145/3597503.3623331}
\acmISBN{979-8-4007-0217-4/24/04}

\begin{document}

%
\title{\toolname: Debugging Block-Based Programs in \Scratch}

%
\author{Adina Deiner}
\affiliation{%
  \institution{University of Passau}
  \country{Germany}
}

\author{Gordon Fraser}
\orcid{0000-0002-4364-6595}
\affiliation{%
  \institution{University of Passau}
  \country{Germany}
}

%

%
\begin{abstract}
  While professional integrated programming environments support
  developers with advanced debugging functionality, block-based
  programming environments for young learners often provide no support
  for debugging at all, thus inhibiting debugging and preventing
  debugging education.
  In this paper we introduce \toolname, an extension of the popular
  block-based programming environment \Scratch that provides the
  missing debugging support. \toolname allows controlling the
  executions of \Scratch programs with classical debugging
  functionality such as stepping and breakpoints, and it is an
  omniscient debugger that also allows reverse stepping. To support
  learners in deriving hypotheses that guide debugging, \toolname is
  an interrogative debugger that enables to ask questions about
  executions and provides answers explaining the behavior in question.
  In order to evaluate \toolname, we survey the opinions of teachers,
  and study the effects on learners in terms of debugging
  effectiveness and efficiency. We find that teachers consider
  \toolname to be useful, and children can use it to debug faulty
  programs effectively. However, systematic debugging requires
  dedicated training, and even when \toolname can provide correct
  answers learners may require further help to comprehend faults and
  necessary fixes, thus calling for further research on improving
  debugging techniques and the information they provide.
\end{abstract}


%
%

\ccsdesc[500]{Software and its engineering~Integrated and visual development environments}
\ccsdesc[500]{Software and its engineering~Software testing and debugging}

%
\keywords{Debugging Tools, Omniscient Debugging, Interrogative Debugging, Scratch, Computer Science Education}

%

%
\maketitle

\begin{acronym}[UMLX]
    \acro{AST}{Abstract Syntax Tree}
    \acro{CDG}{Control Dependence Graph}
    \acro{CFG}{Control Flow Graph}
    \acro{DDG}{Data Dependence Graph}
    \acro{GUI}{Graphical User Interface}
    \acro{SVG}{Scalable Vector Graphic}
    \acro{TAM}{Technology Acceptance Model}
    \acro{VM}{Virtual Machine}
\end{acronym}

\section{Introduction}\label{ch:introduction}

\begin{figure}[t]
  \centering
  \begin{subfigure}{0.50\columnwidth}
    \includegraphics[width=\textwidth]{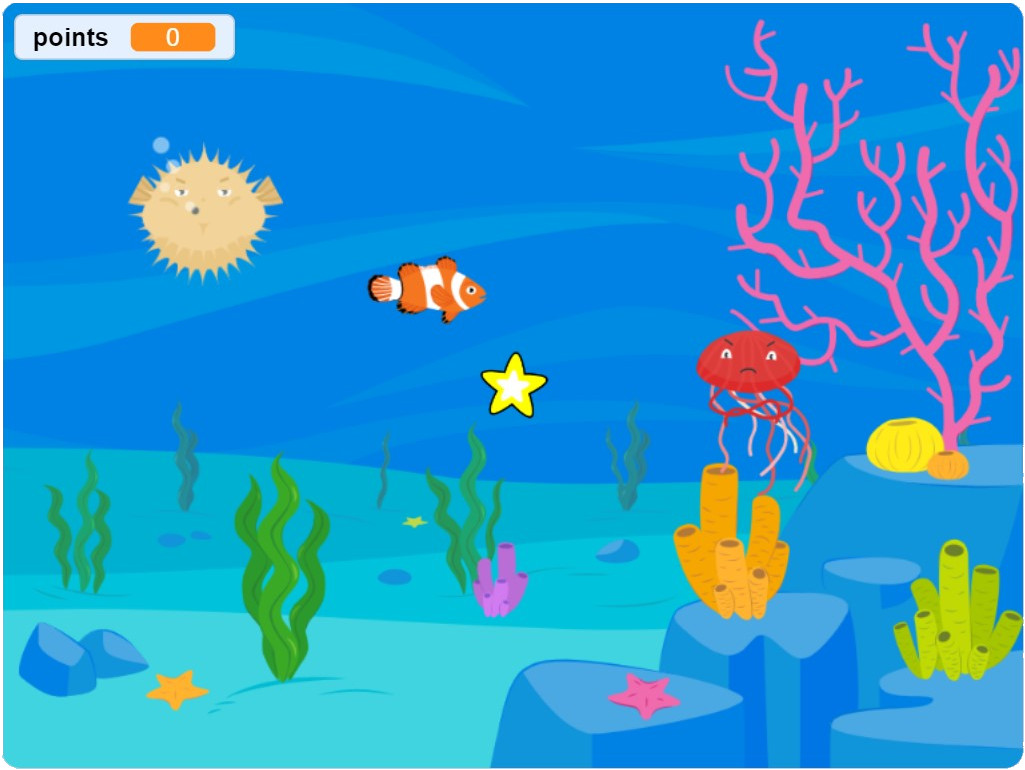}
    \caption{Stage of the program.}
    \label{fig:stage-stars}
  \end{subfigure}
  \hfill
  \begin{subfigure}{0.40\columnwidth}
    \includegraphics[width=\textwidth]{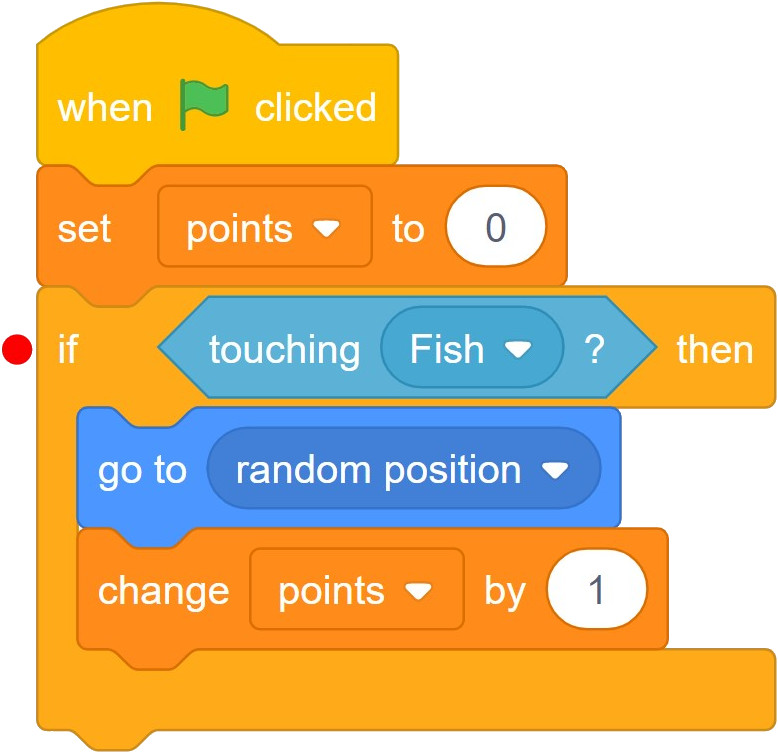}
  \caption{Breakpoint on the block checking if star and fish touch.}
  \label{fig:stars-code-buggy-breakpoint}
\end{subfigure}
\caption{The \Scratch program \textmessage{``Collect the Stars''}, containing a typical fault: The check for whether the fish touches the star should be contained in a loop.}
\end{figure}

Debugging is one of the most time consuming activities during software
development~\cite{maintainingmentalmodels, tassey,
  maintenance}. Debugging is also a frequent activity when programming
with the block-based programming language
\Scratch~\cite{maloney2010scratch}, one of the most popular languages
for introductory
coding.\footnote{\url{https://scratch.mit.edu/statistics}, last
  accessed March 2023} Although the visual composition of blocks
prevents syntactical errors and enables learners to quickly create
games and animations, programs can nevertheless implement wrong
functionality. Finding the cause of a failure can be difficult,
especially for programming
beginners~\cite{fitzgerald,murphy,Jadud2004AFL,beth,vessey} as well as
for teachers trying to support
their learners~\cite{kim2018debugging,tsan2022analysis}.

Consider the simple example \Scratch program \textmessage{``Collect the
  Stars''} shown in \cref{fig:stage-stars}, where the clownfish
\hspace{0.3mm}
\inlineFigure{0.05}{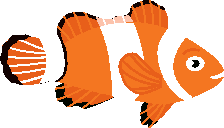}
\hspace{0.3mm} controlled by the user has to collect the stars
\hspace{0mm}
\inlineFigure{0.08}{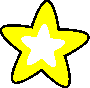}
\hspace{0mm} in the underwater world. Every time the fish touches the
star, the star should change its position and increment the number of
points by one. When executing the program, nothing happens when the
fish touches the star. How can the programmer of this game find out
what caused the failure?

Professional programming environments provide debuggers
facilitating the debugging activity~\cite{zeller}, but \Scratch does
not provide any debugging support.
%
In this paper we therefore introduce \toolname, which contributes
debugging functionality to \Scratch, such as the ability to pause the
execution of a program, or to set breakpoints and execute a program
step by step, even backwards in
time. \Cref{fig:stars-code-buggy-breakpoint} shows a breakpoint on the
block checking if the fish is touching the star. The red dot
visualizes the breakpoint and results in pausing the execution of the
program every time the block is executed, which allows to investigate
the program state used to evaluate the condition. The breakpoint
reveals that the block is only executed once at the start of the
program and the condition is not checked afterwards. The cause of the
failure is found---the if-condition should be contained in a loop, the
omission of which is a common mistake made by
beginners~\cite{fradrich2020common}.

\begin{figure}[t]
  \includegraphics[width=0.8\columnwidth]{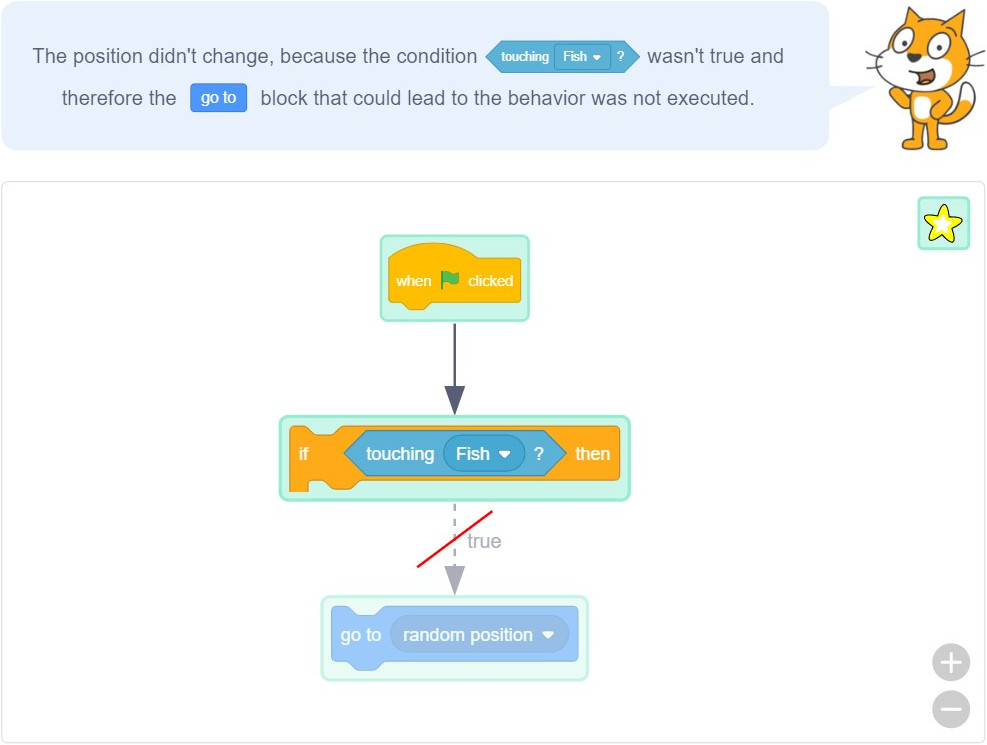}
  \caption{Answer for the question \textmessage{``Why didn't the position of sprite Star change?''} }
  \label{fig:answer-position}
\end{figure}
  
\begin{figure}[t]
  \centering
  \includegraphics[width=0.8\columnwidth]{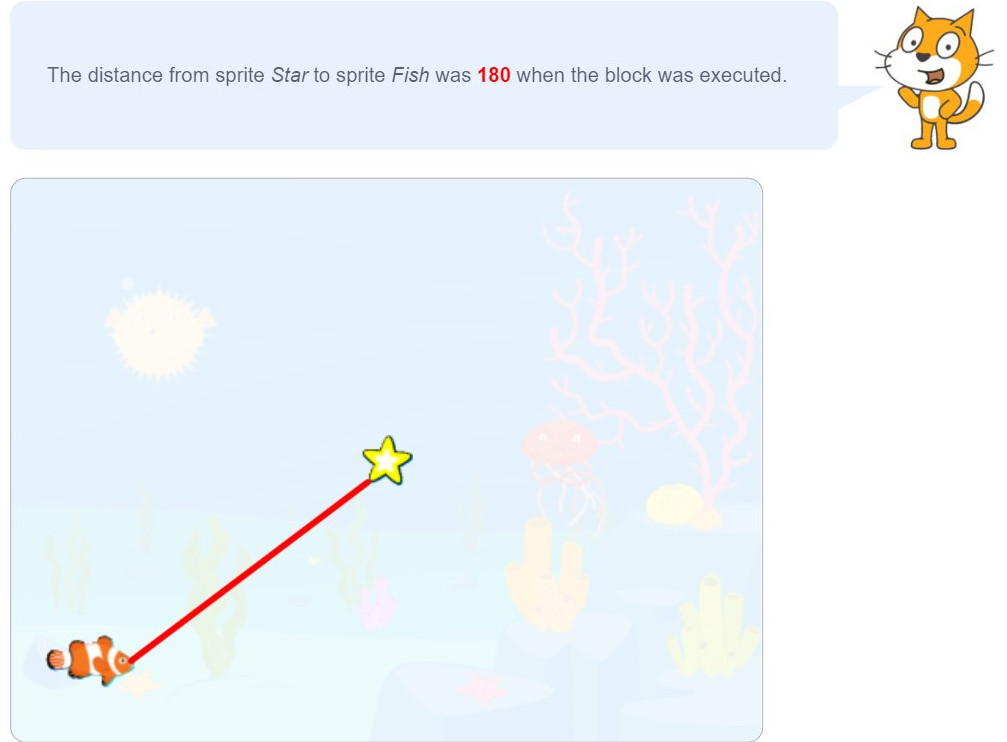}
  \caption{Answer for the question \textmessage{``Why didn't the condition \hspace{0mm} \inlineFigure{0.12}{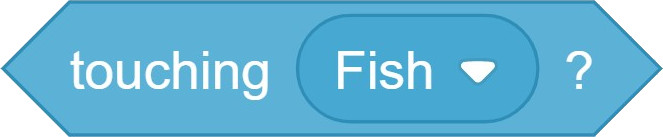} \hspace{0mm} evaluate to true?''} }
  \label{fig:answer-condition}
\end{figure}

Setting the breakpoint requires a hypothesis about the cause of the
failure, but deriving such hypotheses is
difficult~\cite{SoftwareMaintenanceStudy}. Therefore, \toolname
uses interrogative debugging~\cite{AliceWhyline}, which helps
deriving hypotheses by allowing to pose questions one naturally would want to
ask about the execution of a program.
Once a question is selected, the debugger calculates an answer
explaining the behavior and directing the programmer to the fault. For
our example, the debugger makes it possible to ask the question
\textmessage{``Why didn't the position of sprite Star change?''} and
generates the answer visualized in \cref{fig:answer-position}, from
which the programmer can conclude that the condition
\setscratch{scale=0.6}\boolsensing{touching \selectmenu{Fish} ? }
never evaluated to true. Furthermore, it is possible to ask the
question \textmessage{``Why didn't the condition}
\setscratch{scale=0.6}\boolsensing{touching \selectmenu{Fish} ? }
\textmessage{evaluate to true?''} and the debugger provides the answer
shown in \cref{fig:answer-condition}. Since this evaluation to false is the only time the
condition was checked, this helps understanding there is a missing loop.

In detail, the contributions of this paper are as follows:
\begin{itemize}
\item We introduce \toolname, the first debugger for \Scratch, and the
  first approach of omniscient and interrogative debugging for a
  block-based programming environment.
  %
\item We collate questions for \Scratch-like programs, and provide
  novel answer types for block-based programs.
  %
\item We empirically evaluate \toolname using a survey with teachers
  and a controlled study with pupils.
\end{itemize}

Overall, our evaluation demonstrates that \toolname is intuitive and
effective, and brings systematic debugging to the world of block-based
programming. The teachers we survey confirm that \toolname is useful,
and pupils taught debugging are able to fix bugs using \toolname,
often even more effectively so than without debugger.  However, we
also find that maintaining the simplicity of block-based programming is
challenging when presenting textual questions and answers, and young
programming learners need to be explicitly educated on systematic
debugging.


\section{Background}

\subsection{The \Scratch Programming Environment}

\textit{\Scratch} is a block-based programming language primarily
aimed at children and is increasingly used in schools and coding clubs
to introduce programming~\cite{maloney2010scratch}. Program statements
in \Scratch are \emph{blocks} that can be stacked to create
\emph{scripts}. Blocks can also contain holes of different shapes into
which other blocks of matching shapes can be placed, for example to
report the values of attributes or variables (\emph{reporter blocks})
or conditions (\emph{Boolean blocks}). Block shapes ensure that
resulting programs are syntactically
valid~\cite{ScratchProgrammingForAll}.

\Scratch programs are structured into \emph{targets}: The
\textit{stage} is the application window containing the background
image, and \textit{sprites} are rendered on top of the stage.
Each target contains scripts that control its behavior. The first
block of a script is always an event handler (\emph{hat block}), and
the execution of a script is triggered every time the event
represented by the hat block occurs. One particular such event is the
green flag
\includegraphics[height=0.8\baselineskip]{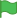},
which the user can click to start the program execution. Scripts are
executed in separate threads by the \Scratch \acf{VM}, and thread
switching occurs when scripts encounter a waiting state, reach the end
of a loop, or have no more blocks to execute.  To avoid that learners
have to deal with confusing error messages, \Scratch uses a `failsoft'
approach where runtime errors are swallowed and execution is resumed
without informing the user~\cite{hromkovic2021problem}.

\begin{figure}[t]
   \centering
   \includegraphics[width=\columnwidth]{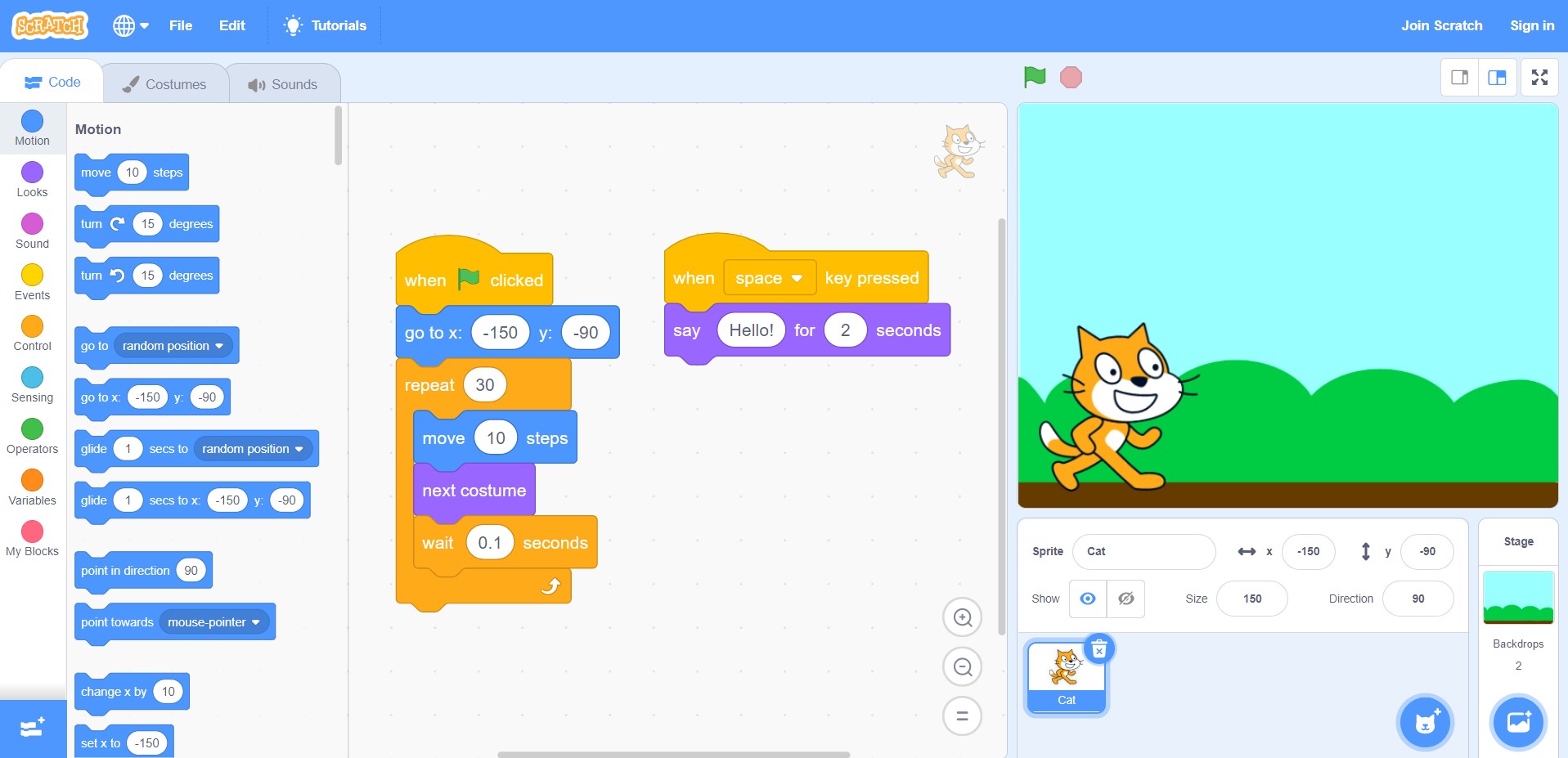}
   \caption{The \Scratch \acs{GUI} consisting of a block palette containing available blocks on the left, the cat's source code in the center, the stage on the upper right and the target editor on the bottom right.}
   \label{fig:original-scratch}
 \end{figure}

The \Scratch's \acf{GUI} is divided into four main sections
(\cref{fig:original-scratch}): The \textit{block palette} on the left
contains all available blocks divided into different color-coded
categories. Blocks can be dragged out of the block palette and then
dropped and snapped together in the \emph{coding area} in the middle
to define the behavior of the currently selected target. The upper
right contains the \emph{stage}, which renders the programmed
behavior. The bottom right area enables selecting and editing targets.

\subsection{Debugging Tools}
\label{sec:debuggers}

\textit{Debugging} is the activity of detecting, locating and
eliminating program errors~\cite{iso_debug}.
%
A typical \textit{debugger} provides three
functionalities~\cite{zeller}:
First, it supports tracing by pausing the execution of a program at
any point in time or on specified conditions. The most common
available pausing condition is a \textit{breakpoint}, which is a
specific location in the program or condition on the execution that
results in pausing the execution when reached.
%
Second, once the execution is paused, classical debuggers provide
context information, such as the values of variables or the current
stack trace.
%
Third, it allows resuming the execution, either until the next pause,
or the next statement, which enables stepping through a program.

\label{sec:omniscient-debugger}
One of the reasons why debugging is difficult is the temporal or
spatial chasm between the cause and the symptom of an
error~\cite{eisenstadt}. Although programs are executed forward in
time, debugging requires thinking backwards from the failure to
discover the cause.
\textit{Omniscient debuggers}~\cite{balzer1969exdams,barr2016time}
record executions to enable users to explore the execution
history~\cite{lewis2003debugging}, going back and forth to arbitrary
moments instead of having to restart programs multiple
times~\cite{pothier2009back}.

Debugging is triggered by a question about the behavior and methods
like breakpoints or code stepping require coming up with a hypothesis
about the cause of the behavior, which may be
difficult~\cite{JavaWhyline2}. False assumptions may lead to a
time-consuming investigation of unrelated code and in the worst case
prevent detecting the error. Ko and Myers observed that programmers
naturally would like to ask \WhyDid questions about unexpected output
that did occur and \WhyDidnt questions about expected output that did
not occur~\cite{AliceWhyline}. \emph{Interrogative debuggers} such as
\whyline for \alice~\cite{AliceWhyline} and \java~\cite{JavaWhyline2}
allow to directly ask such questions via a ``why''-menu, in which
\WhyDid and \WhyDidnt questions can be selected for available objects.
Answers point out the cause of the queried behavior as graphs
explaining causality in terms of data and control flow, derived from
execution traces, dynamic slices, and code analysis.

\subsection{Debugging in Block-Based Programming}
\label{sec:bb_debugger}

Since block-based programming environments generally follow a
preventive approach that avoids exposing learners to
errors~\cite{hromkovic2021problem}, debugging tools are usually
missing. \Scratch offers no support to trace executions, and to
inspect states one needs to resort to
{\setscratch{scale=0.4}\begin{scratch} \blocklook{say
      \ovalnum{}}\end{scratch}} blocks to have sprites communicate
values at runtime.
A recent survey of block-based programming
environments~\cite{selwyn2022blocks} confirms there are only few
exceptions: \snap~\cite{harvey2010bringing} allows pausing executions
in addition to starting and stopping the entire execution, and
Microsoft's \makecode~\cite{ball2019microsoft} supports
breakpoints. An experimental extension of the hybrid (text and
block-based) \pencilcode~\cite{bau2015pencil} programming environment
highlights executed blocks using arrows and supports step-wise
execution and variable tracking~\cite{boss2015visual}. Stepping,
breakpoints and watches are also part of a proposed visual
debugger~\cite{DBLP:conf/ihsed/SavidisS19} for Google's
\blockly\footnote{https://developers.google.com/blockly}.
\alice~\cite{cooper2000alice}, which is not strictly block-based but
similarly constrains the syntactic validity of programs,
provides no dedicated debugging support, but the \whyline
interrogative debugger was originally implemented for
\alice~\cite{AliceWhyline}.
The general lack of debugging support in block-based 
environments is concerning: A recent study~\cite{martinez2020novice}
found that students found and fixed fewer bugs in block-based
environments than in hybrid
ones. 


There is a growing awareness of the importance of teaching
debugging~\cite{michaeli2019improving} and the lack of debugging in
education~\cite{michaeli2019current}. Several studies on the processes
applied by young learners~\cite{tenorio2019debugging,yan2021exploring}
in block-based programming revealed that they tend to debug
unconsciously without a systematic but with a tinkering-based
approach~\cite{dong2019defining}.
To counter this, research has resulted in instructional 
materials~\cite{chiu2015guided,ahn2017supporting}, learning
trajectories~\cite{rich2019k}, teaching
strategies~\cite{gao2023flipped,socratous2021structured}, serious
games~\cite{lee2014principles,miljanovic2017robobug,luxton2018ladebug},
and unplugged~\cite{ahn2022unplugged} and
tangible~\cite{sipitakiat2012robo} environments for teaching debugging
of block-based programs. Debugging tools, however, are generally not
included.
Debugging tools would furthermore not only be important for learners,
but their educators also require skills and confidence in debugging to
teach how to debug block-based programs, and to support students in
the classroom. While teachers have been reported to struggle finding
and fixing bugs in block-based
programs~\cite{kim2018debugging,tsan2022analysis}, tools have been
shown to be helpful particularly for
teachers~\cite{greifenstein2021effects}.




\section{The \toolname~\Scratch Debugger}

\label{sec:debugger}

\toolname is the first debugger for \Scratch, and provides all the
debugging functionality described in \cref{sec:debuggers}. In order to
provide this debugging functionality, \toolname extends the \Scratch
\acs{VM} to trace executions (\cref{sec:tracing}), which results in
collecting all the information that is necessary for
debugging~\cite{rosenberg}. \toolname also extends the \Scratch user
interface for controlling the debugger, allowing users to pause
executions, set breakpoints, or step through program executions
forward and backward (\cref{sec:control}). Execution traces are also
prerequisite for creating questions and answers during debugger
interrogation (\cref{sec:interrogation}).

\subsection{Tracing Executions}
\label{sec:tracing}

\toolname instruments the \acs{VM} with a custom tracer, which records
which blocks have been executed in which order. Every time a block is
executed by the \acs{VM}, the tracer appends a new trace entry to a
trace, storing the executed block and the program state resulting from
the execution of the block.
The trace is automatically cleared every time the green flag is
clicked, because then a new program run starts, and the trace is also
reset whenever the code is changed. 

A trace entry records the execution of a block and the resulting
program state. It stores (1) the identifier of the block,
(2) the type of the block, (3) the names of input parameters and (4)
variables, (5) all evaluated parameter values, as well as (6) a list
of trace entries recording the execution of reporter and Boolean
blocks used by the block.
For non-parameter blocks the trace entry also stores (7) the execution
time of the block, and (8) the id of the target instance executing the
block. Finally, the trace entry records (9) the program state after
the execution of the block. For each target, this state consists of
the target's costume or backdrop, visual effects, volume, sound state
and variable values. If the target is a sprite, it also contains the
position, direction, rotation style, size, visibility, draggability,
layer and state of its speech or think bubble. If the executed block
contains a \setscratch{scale=0.6}\boolsensing{touching
  \selectmenu{mouse-pointer} ? } condition, the tracer also saves the
current position of the mouse-pointer.
Finally, to enable reverse execution the execution state information
includes the currently active threads, the id of the next thread to
execute, the number of executed steps and the last executed blocks.

To reduce the size of execution traces and the computation time of
their analysis, the tracer applies two optimizations. First, \Scratch
allows displaying the values of variables or attributes on the stage,
which internally is treated like block executions. Since these have no
effects on the state, the tracer ignores them. Second, blocks that
halt the execution of their thread waiting for a timeout or condition
are internally processed like repeated block executions in the
\acs{VM}. For such blocks ({\setscratch{scale=0.4}\begin{scratch}\blockmove{glide \ovalnum{time} seconds to ...}\end{scratch}, \begin{scratch} \blocklook{say
  \ovalnum{message} for \ovalnum{time} seconds}\end{scratch}, \begin{scratch} \blocklook{think
  \ovalnum{message} for \ovalnum{time} seconds}\end{scratch}, \begin{scratch} \blocksound{play
  sound \selectmenu{sound} until done}\end{scratch}, \begin{scratch} \blockcontrol{wait
  \ovalnum{time} seconds}\end{scratch}, and \begin{scratch} \blockcontrol{wait until
  \boolsensing{condition}}\end{scratch}}), only the initial and final execution states
are stored.

The tracer can be enabled or disabled by the user
with a new button in the control panel above the stage. If tracing is
active the button has a blue background color
\inlineFigure{0.1}{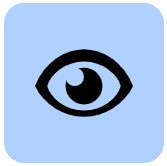} and otherwise it has a
transparent background color
\inlineFigure{0.1}{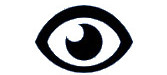}. Toggling the value
also clears the trace to avoid gaps between trace entries.

\subsection{Execution Control}
\label{sec:control}

\begin{figure}[tb]
  \centering
  \begin{subfigure}{0.3\columnwidth}
    \includegraphics[height=0.9cm]{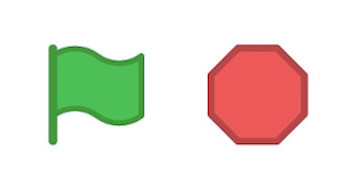}
  \caption{Original.}
  \label{fig:original-control-panel}
\end{subfigure}
  \hspace{0.1cm}
  \begin{subfigure}{0.6\columnwidth}
  \includegraphics[height=1cm]{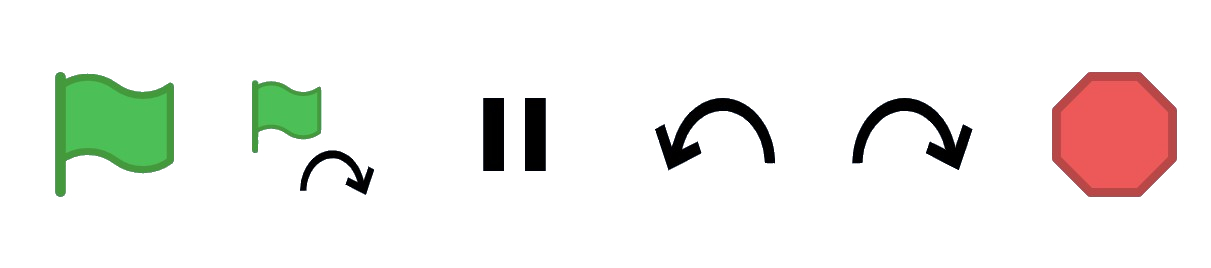}
  \caption{Extended by \toolname.}
  \label{fig:extended-control-panel}
\end{subfigure}
\vspace{-1em}
\caption{\Scratch control panel.}
\vspace{-1em}
%
\end{figure}

\Scratch allows starting and stopping executions using the control
panel above the stage shown in
\cref{fig:original-control-panel}. \toolname extends this with
debugging functionality (\cref{fig:extended-control-panel}).
In particular, \toolname provides
functionality to pause an execution and to resume it again once paused. Moreover, we enable the user to execute a
program step by step and to rewind the execution to any traced program
state. In addition, it is possible to add breakpoints to blocks such
that the program pauses every time the block would be executed.

\subsubsection{Pausing}

\begin{figure}[h]
  \centering
  \includegraphics[width=\columnwidth]{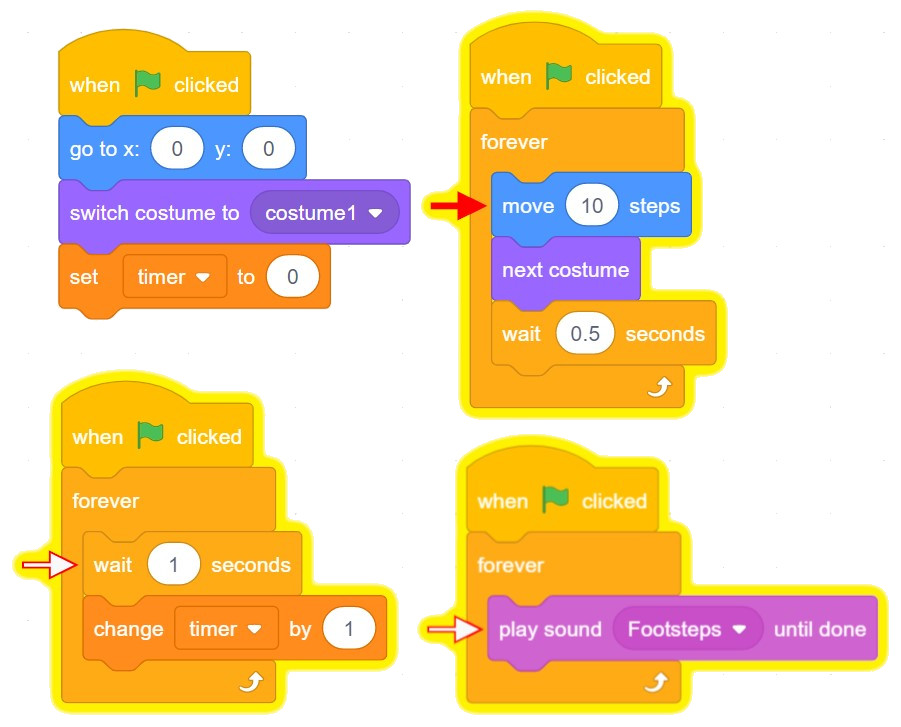}
  \vspace{-2em}
  \caption{Arrows emphasizing the last executed block of each active script and the overall last executed block.}
  \vspace{-1em}
  \label{fig:block-arrows}
\end{figure}

The \textmessage{``Pause / Resume''} button
(\inlineFigure{0.4}{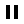} if the
execution can be paused, and
\inlineFigure{0.4}{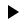} if the
execution can be resumed) allows halting executions. If the program is
not running, 
the button is disabled.
%
When the execution is paused, a white arrow with a red border is shown
next to the last executed block of each active script, and a red arrow
is shown next to the overall last executed block
(\cref{fig:block-arrows}). In addition, the selected target is changed
to the one containing the overall last executed block, which is
blinked. 

\subsubsection{Stepping}

\begin{figure}[h]
  \centering
  \subfloat[Before.]{\includegraphics[width=0.22\columnwidth]{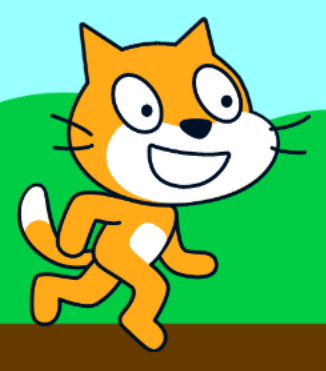}}
  \hfill
  \subfloat[During.]{\includegraphics[width=0.22\columnwidth]{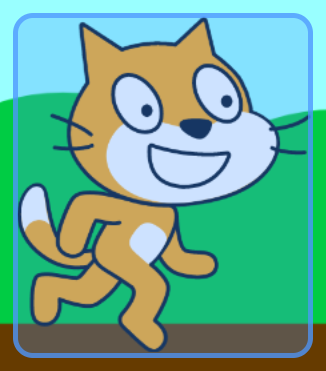}}
  \hfill
  \subfloat[After.]{\includegraphics[width=0.22\columnwidth]{figures/execution-control/executed-target-without-border}}
  \hfill
  \vspace{-1em}
  \caption{Highlighted target instance during the execution of a block.}
  \vspace{-1em}
  \label{fig:target-highlight}
\end{figure}

If the execution is paused, the user can execute a single block at a
time by clicking the \textmessage{``Step Over''} button \hspace{0.1mm}
\inlineFigure{0.0024}{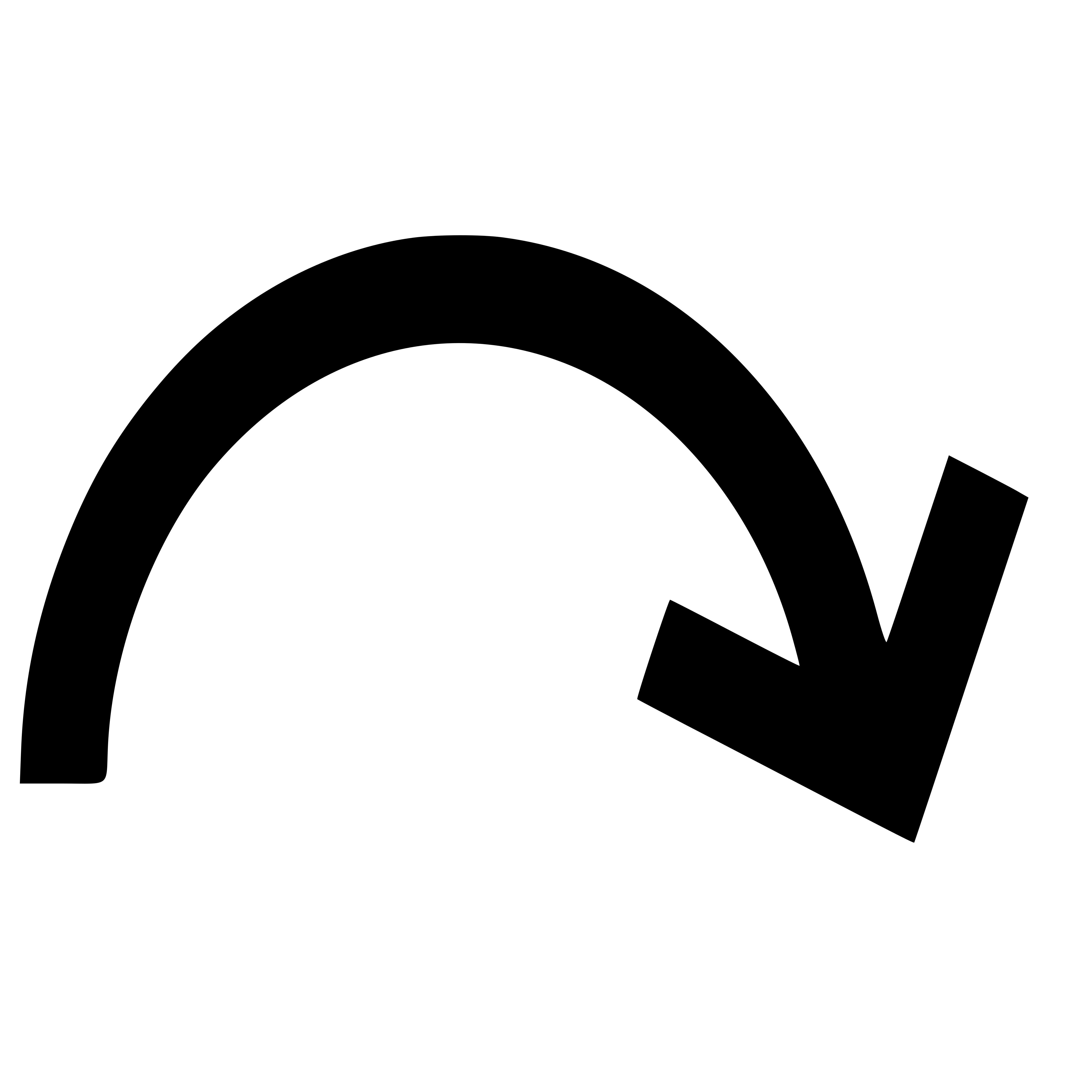}
\hspace{0.1mm} in the control panel
(\cref{fig:extended-control-panel}), which
informs the \Scratch \acs{VM}'s runtime about stepping, such that 
the modified \Scratch
pauses
after one executed block, adhering to the thread execution order.
During the execution of a block, the target instance executing the
block is highlighted with a blue overlay
 (\cref{fig:target-highlight}).
This is especially useful when multiple
instances of the same target (i.e., \emph{clones}) exist and it is
unclear which one executed the block.  While stepping, the arrows
emphasizing the last executed blocks are shown.
%
%
%
Finally, the
\inlineFigure{0.0024}{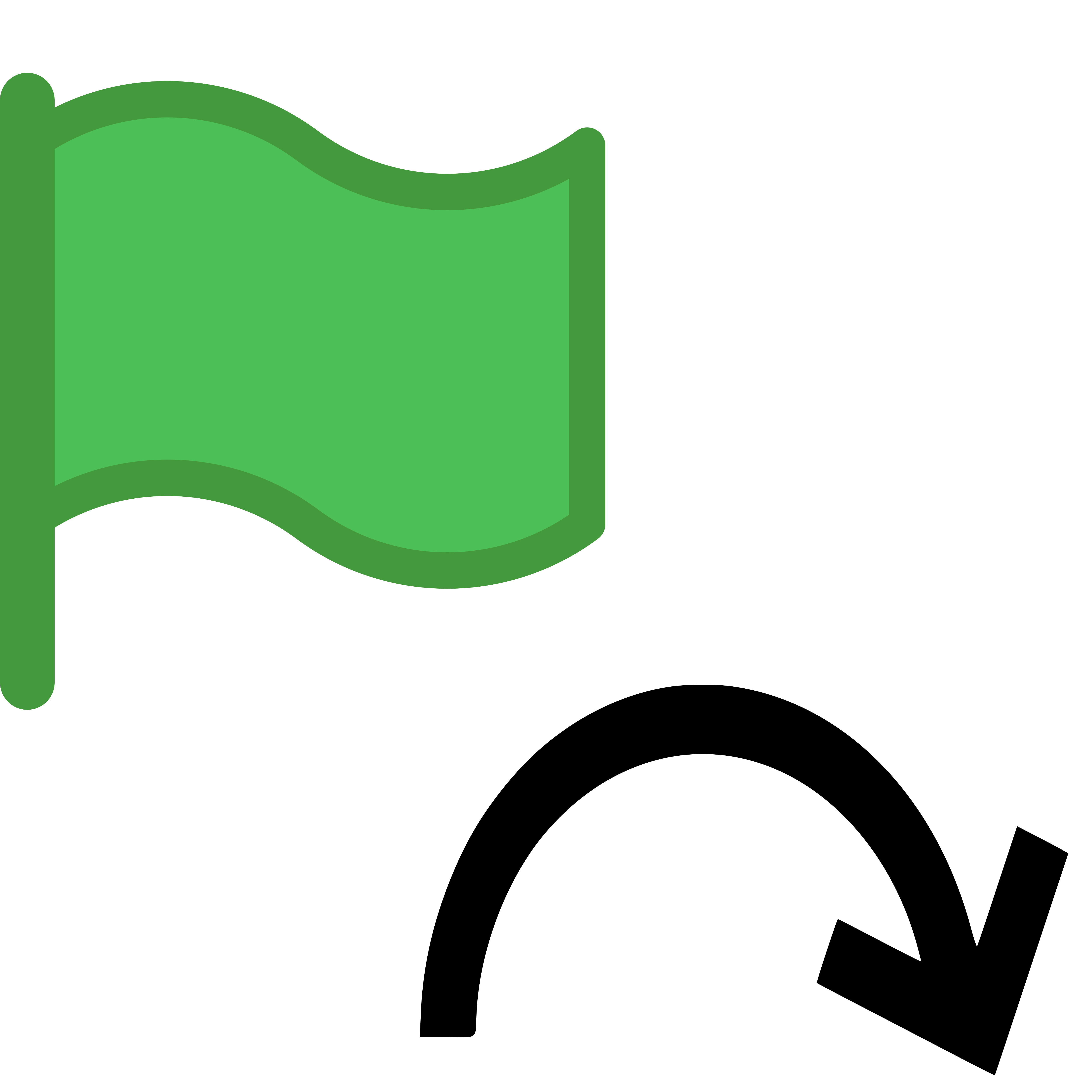}
button (\cref{fig:extended-control-panel}) allows executing the first
block of a program, by activating stepping and then invoking the
program like when the green flag is clicked. This causes the runtime
to start the program execution, execute the first block and then pause.

\subsubsection{Reverse Execution}

\begin{figure}[t]
  \centering
  \subfloat[Final state.]{\includegraphics[width=0.3\columnwidth]{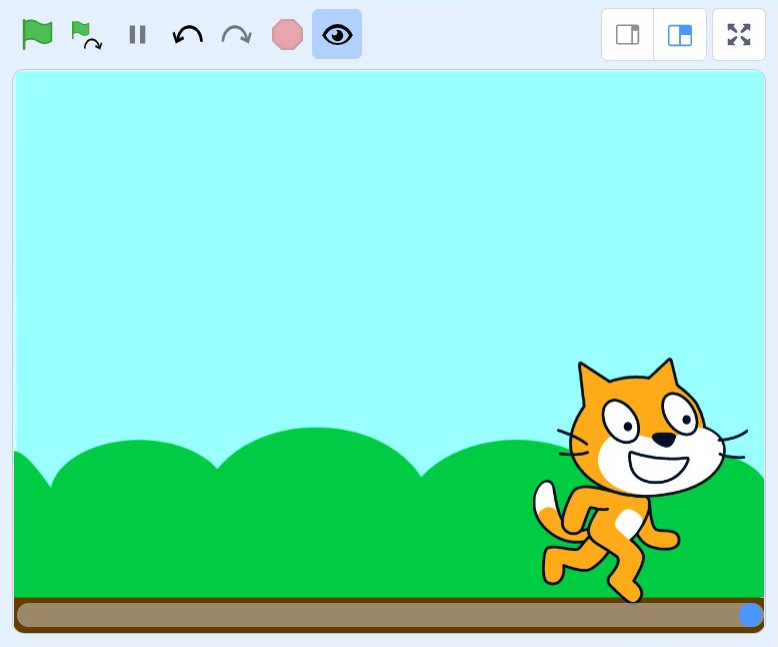}\label{fig:rewound_execution-1}}
  \hfill
  \subfloat[Intermediate state.]{\includegraphics[width=0.3\columnwidth]{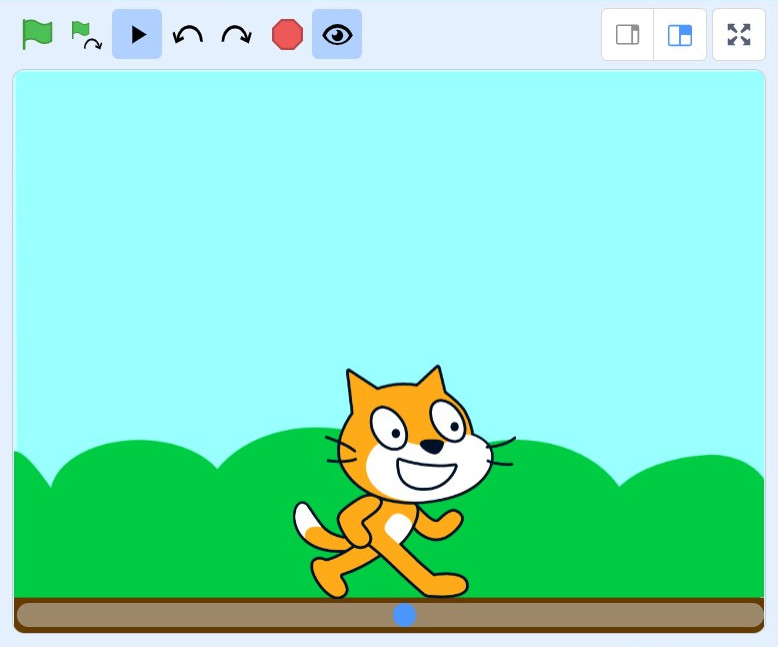}\label{fig:rewound_execution-2}}
  \hfill
  \subfloat[Initial state.]{\includegraphics[width=0.3\columnwidth]{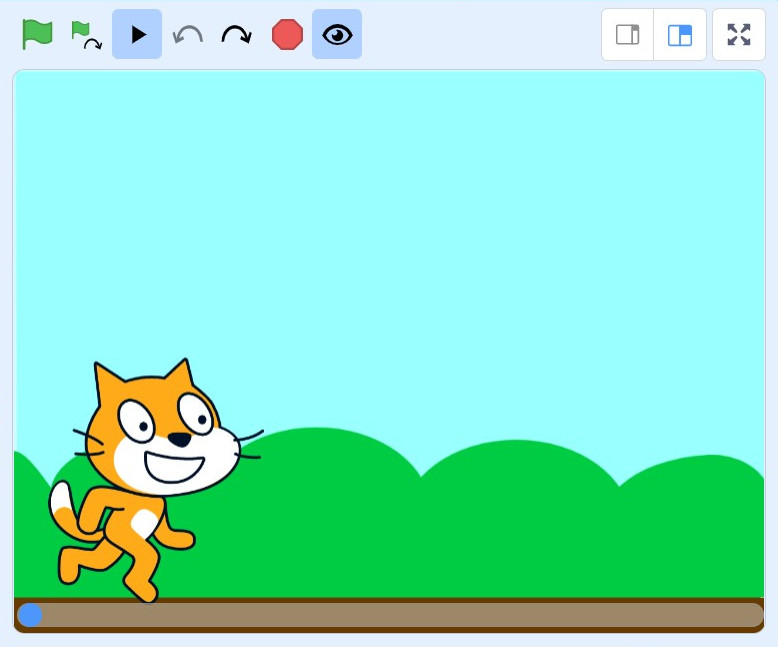}\label{fig:rewound_execution-3}}
  \hfill
  \vspace{-1em}
  \caption{Execution states during rewinding of an execution.}
  \label{fig:rewound_execution}
\end{figure}

In order to jump to arbitrary points in a trace, the user interface
provides a slider to select the trace position (see bottom of
\cref{fig:rewound_execution}).
In addition, a \textmessage{``Step back''} button
\hspace{0.1mm}
\inlineFigure{0.0024}{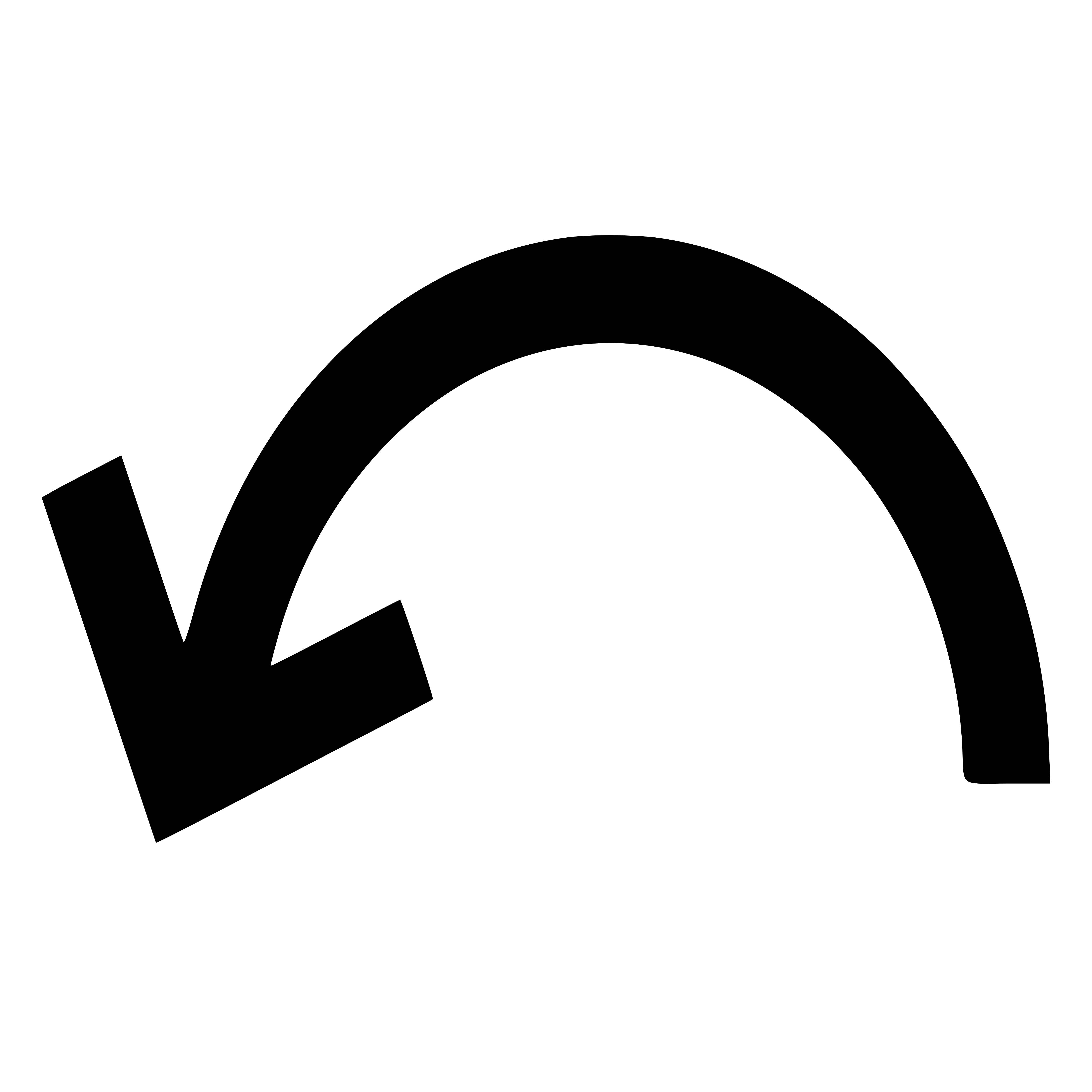}
\hspace{0.1mm} above the stage allows to rewind the execution by one
executed block (if there is a predecessor).
%
%
In either case the recorded program state is restored in terms of the clones existing in that state, attributes of all target instances, and thread status.
%
%
If the selected trace entry is not the last element in the trace,
%
the \textmessage{``Step over''} button does not execute the next
block, but instead increments the index of the selected trace entry by
one and restores the state of this trace entry. Clicking the
\textmessage{``Resume''} button does not resume the execution as
usual, but deletes all trace entries succeeding the selected one and
resumes the execution at the corresponding program state.


\subsubsection{Breakpoints}

\begin{figure}[tb]
  \centering
  \subfloat[Unset breakpoint.]{\includegraphics[height=0.12\columnwidth]{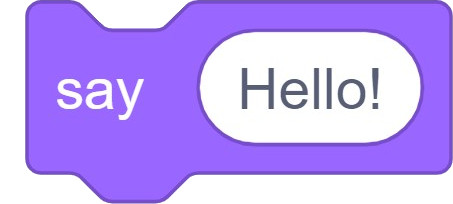}\label{fig:without-breakpoint}}
  \hfill
  \subfloat[Hovering over a breakpoint.]{\includegraphics[height=0.12\columnwidth]{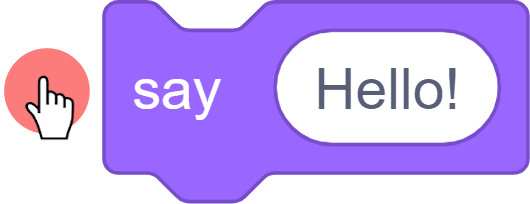}\label{fig:hover-breakpoint}}
  \hfill
  \subfloat[Set breakpoint.]{\includegraphics[height=0.12\columnwidth]{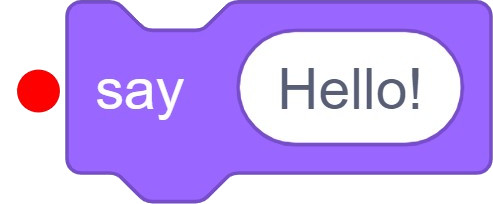}\label{fig:with-breakpoint}}
  \hfill
  \vspace{-1em}
  \caption{Possible states of a breakpoint.}
  \vspace{-1em}
  \label{fig:breakpoint}
\end{figure}

\toolname provides functionality to set a breakpoint on a block, which
results in halting the execution every time it is
executed. Breakpoints are supported by all blocks except reporter,
Boolean and hat blocks. Reporter and Boolean blocks are
implicitly evaluated when the block they are
contained in is executed. Hat blocks
would
interfere with the event triggering the hat block.

Each block stores whether its breakpoint is set. By default, a block's
breakpoint is not set and the block is rendered as usual
(\cref{fig:without-breakpoint}). While hovering the mouse pointer over
the left margin of a block, a red dot visualizing the breakpoint
appears (\cref{fig:hover-breakpoint}). Clicking on it sets
the breakpoint, which leads to the red dot being shown
(\cref{fig:with-breakpoint}) permanently. Another click 
unsets the breakpoint and removes the red dot again. It is also
possible to toggle the breakpoint by clicking the context menu option
\textmessage{``Add Breakpoint''} or \textmessage{``Remove
  Breakpoint''}. 
In order to pause executions at the correct time, the \acs{VM} is
instrumented to check for activated breakpoints before executing
blocks.
%
If a breakpoint is encountered, the execution is paused and the
breakpoint's block is highlighted in red.
%
%
%
%



\section{\Scratch Interrogative Debugging}
\label{sec:interrogation}

\begin{figure}[h]
  \centering
  \includegraphics[width=0.85\columnwidth]{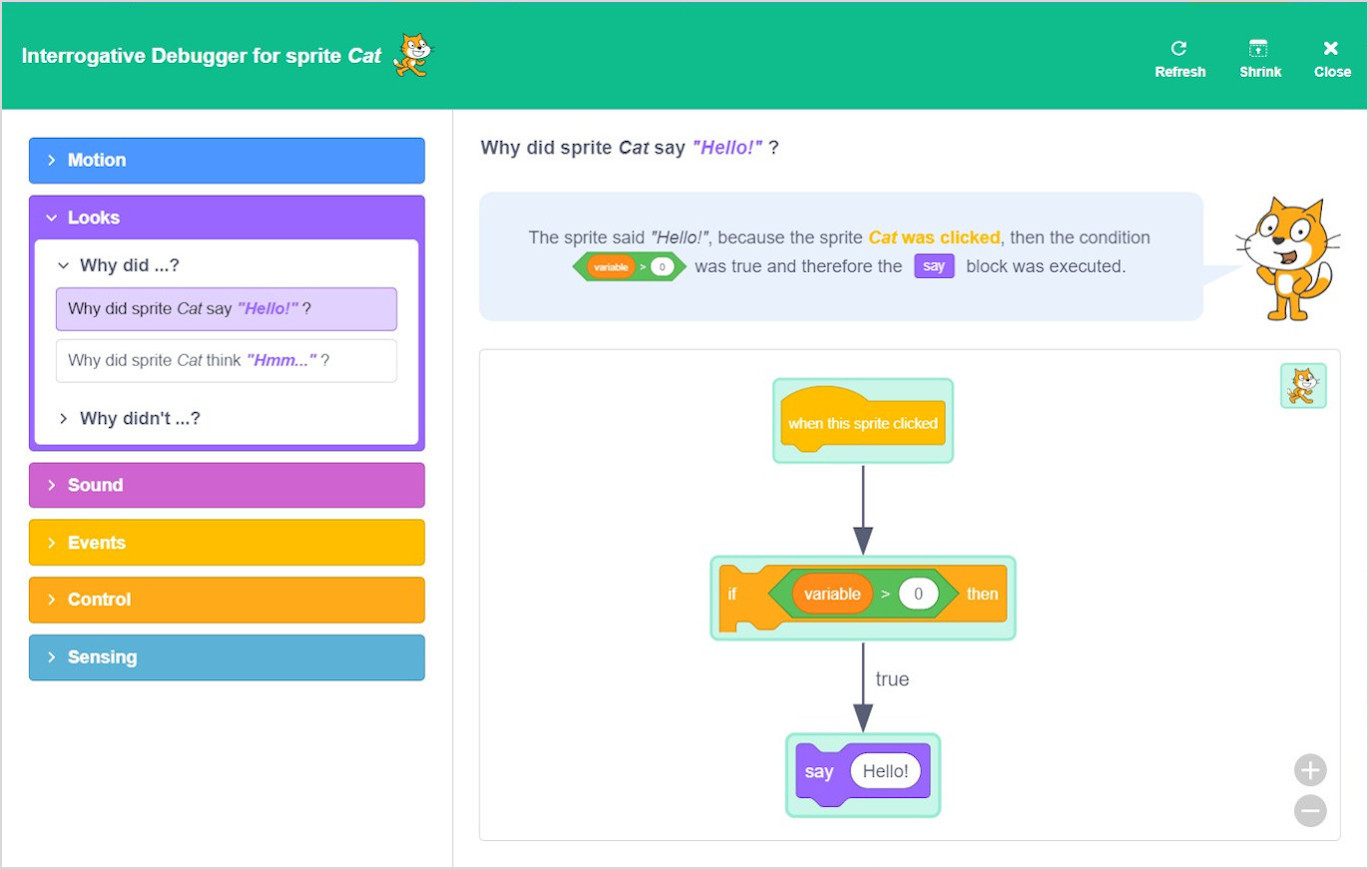}
  \vspace{-0.5em}
  \caption{The dialog of the interrogative debugger.}
  \vspace{-1em}
  \label{fig:dialog}
\end{figure}

In addition to regular and omniscient debugging functionality
(\cref{sec:debugger}), \toolname implements interrogative debugging.
It generates questions the user might want to ask about the execution
of a program
(\cref{sec:questions}), as well as visual and textual answers
for these (\cref{sec:answers}). \Cref{fig:dialog} shows the
full interrogative debugger in action, with categorized questions to
the left, textual answer on the top, and an answer visualized as a
graph beneath.

The principle of interrogative debugging
was originally
introduced in \whyline~\cite{AliceWhyline}.
\toolname is inspired by \whyline, but
differs in a number of ways: First, the user interface for asking
questions is different and designed specifically to blend in
seamlessly in the \Scratch UI, and allows navigation of a new set of
questions tailored specifically for \Scratch programs and
blocks. Second, \toolname uses analyses tailored for \Scratch and a
different approach to represent answers, based on custom answer graphs
as well as visualizations of the \Scratch stage.

\setscratch{scale=0.5}

\subsection{Questions}
\label{sec:questions}

Like \whyline~\cite{AliceWhyline}, at the highest level we distinguish
between positive (\WhyDid) and negative (\WhyDidnt) questions, but
also add a third category (\WhenDid).  \toolname then provides
comprehensive questions covering all attributes (e.g., size, rotation,
position, \ldots) and behaviors of sprites (e.g., movement, rotation,
saying, \ldots), as well as control-flow questions (e.g., why blocks
were or were not executed) and data-flow questions (e.g., regarding
the values of attributes, variables, or conditions).

Questions are generated based on the code of the program that is being
debugged, such that only questions about relevant hypotheses are
shown.  To further avoid overwhelming users with too many questions,
they first have to select an object of interest, which can be a target
or a block. For each possible object the context menu is extended with
the option \textmessage{``open questions''}, which pauses the
execution if running, opens the interrogative debugger and triggers
the generation of questions related to the selected object.
The questions are categorized according to the category of
the corresponding block:
\textcolor[HTML]{4C97FF}{\textsc{Motion}}, 
\textcolor[HTML]{9966FF}{\textsc{Looks}}, 
\textcolor[HTML]{C94FC9}{\textsc{Sound}}, 
\textcolor[HTML]{FFBF00}{\textsc{Events}}, 
\textcolor[HTML]{FFAB19}{\textsc{Control}}, 
\textcolor[HTML]{5CB1D6}{\textsc{Sensing}}, 
\textcolor[HTML]{59C059}{\textsc{Operators}}, 
\textcolor[HTML]{FF8C1A}{\textsc{Variables}} and
\textcolor[HTML]{FF661A}{\textsc{Lists}}.
Additionally, 
\textcolor[HTML]{D0112A}{\textsc{Execution}} contains general
questions about the execution of a block. Within each category
questions are further grouped into \WhyDid, \WhyDidnt and \WhenDid
questions, resulting in the hierarchy
located in the left of the dialog
(\cref{fig:dialog}).



\begin{table}[tb]
\vspace{-1em}
\caption{Target questions.}
\vspace{-1em}
\myfloatalign
\begin{tabularx}{\columnwidth}{X}
\toprule
\textsc{\textcolor[HTML]{4C97FF}{Motion}}   \\ 
\textmessagetable{Why did{\color{lightgray}{n't}} the position of <sprite> change?} \\
\textmessagetable{Why did <sprite> move right- / left- / up- / down-wards?}\\
\textmessagetable{Why did{\color{lightgray}{n't}} the direction of <sprite> change?} \\
\textmessagetable{Why did{\color{lightgray}{n't}} the direction of <sprite> change to <direction>°?} \\
\textmessagetable{Why did{\color{lightgray}{n't}} <sprite> point towards <object>?}\\
\textmessagetable{Why did{\color{lightgray}{n't}} <sprite> turn clockwise / counterclockwise?} \\
\textmessagetable{Why did{\color{lightgray}{n't}} <sprite> turn to the right / left?} \\ 
\midrule
\textsc{\textcolor[HTML]{9966FF}{Looks}} \\ 
\textmessagetable{Why did{\color{lightgray}{n't}} <sprite> say / think <message>?} \\
\textmessagetable{Why did{\color{lightgray}{n't}} the size of <sprite> change?} \\
\textmessagetable{Why did the size of <sprite> increase / decrease?} \\
\textmessagetable{Why did{\color{lightgray}{n't}} <sprite> show / hide itself?} \\
\textmessagetable{Why did{\color{lightgray}{n't}} the costume of <sprite> change?} \\
\textmessagetable{Why did{\color{lightgray}{n't}} the costume of <sprite> change to <costume>?} \\
\textmessagetable{Why did{\color{lightgray}{n't}} <sprite> change the backdrop?} \\
\textmessagetable{Why did{\color{lightgray}{n't}} <sprite> change the backdrop to <backdrop>?} \\
\textmessagetable{Why did{\color{lightgray}{n't}} the backdrop change?} \\
\textmessagetable{Why did{\color{lightgray}{n't}} the backdrop change to <backdrop>?} \\ 
\midrule
\textsc{\textcolor[HTML]{C94FC9}{Sound}} \\ 
\textmessagetable{Why did{\color{lightgray}{n't}} <target> play sound <sound>?} \\
\textmessagetable{Why did{\color{lightgray}{n't}} <target> stop all sounds?}  \\
\midrule
\textsc{\textcolor[HTML]{FFBF00}{Events}} \\ 
\textmessagetable{Why did{\color{lightgray}{n't}} <target> broadcast the message <message>?} \\
\textmessagetable{Why did{\color{lightgray}{n't}} <target>  receive the message <message>?} \\ 
\midrule
\textsc{\textcolor[HTML]{FFAB19}{Control}}  \\ 
\textmessagetable{Why did{\color{lightgray}{n't}} <sprite> start as a clone?} \\
\textmessagetable{Why did{\color{lightgray}{n't}} <target> create a clone of <sprite>?}  \\ 
\midrule
\textsc{\textcolor[HTML]{5CB1D6}{Sensing}}  \\ 
\textmessagetable{Why did{\color{lightgray}{n't}} <target> ask <message>?}  \\
\bottomrule
\end{tabularx}
\label{tab:target-questions}
\vspace{-1em}
\end{table}

\subsubsection{Target behavior questions}\label{sec:target-questions}

\begin{figure}[htb!]
  \centering
  \subfloat[Context menu of a sprite.]{\includegraphics[width=0.4\columnwidth]{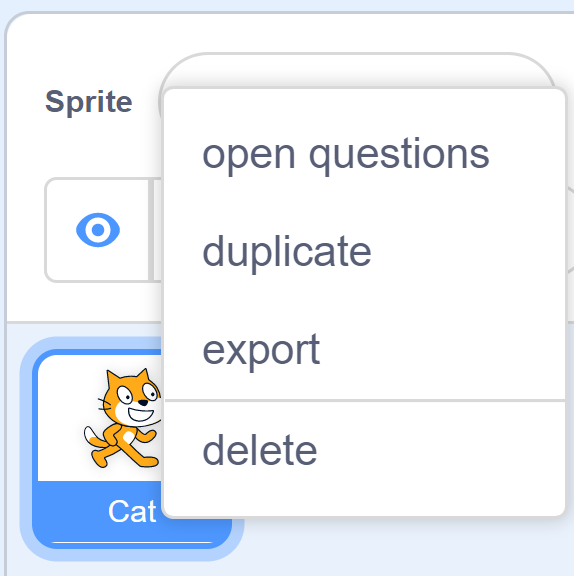}}
  \hspace{11mm}
  \subfloat[Context menu of the stage.]{\includegraphics[width=0.4\columnwidth]{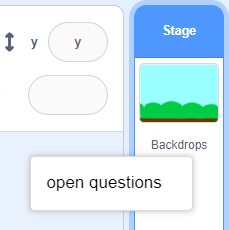}}
  \vspace{-0.5em}
  \caption{The new context menu option \textmessage{``open questions''}.}
  \vspace{-1em}
  \label{fig:open-target-questions}
\end{figure}

\begin{figure}[htb!]
  \centering
  \includegraphics[width=\columnwidth]{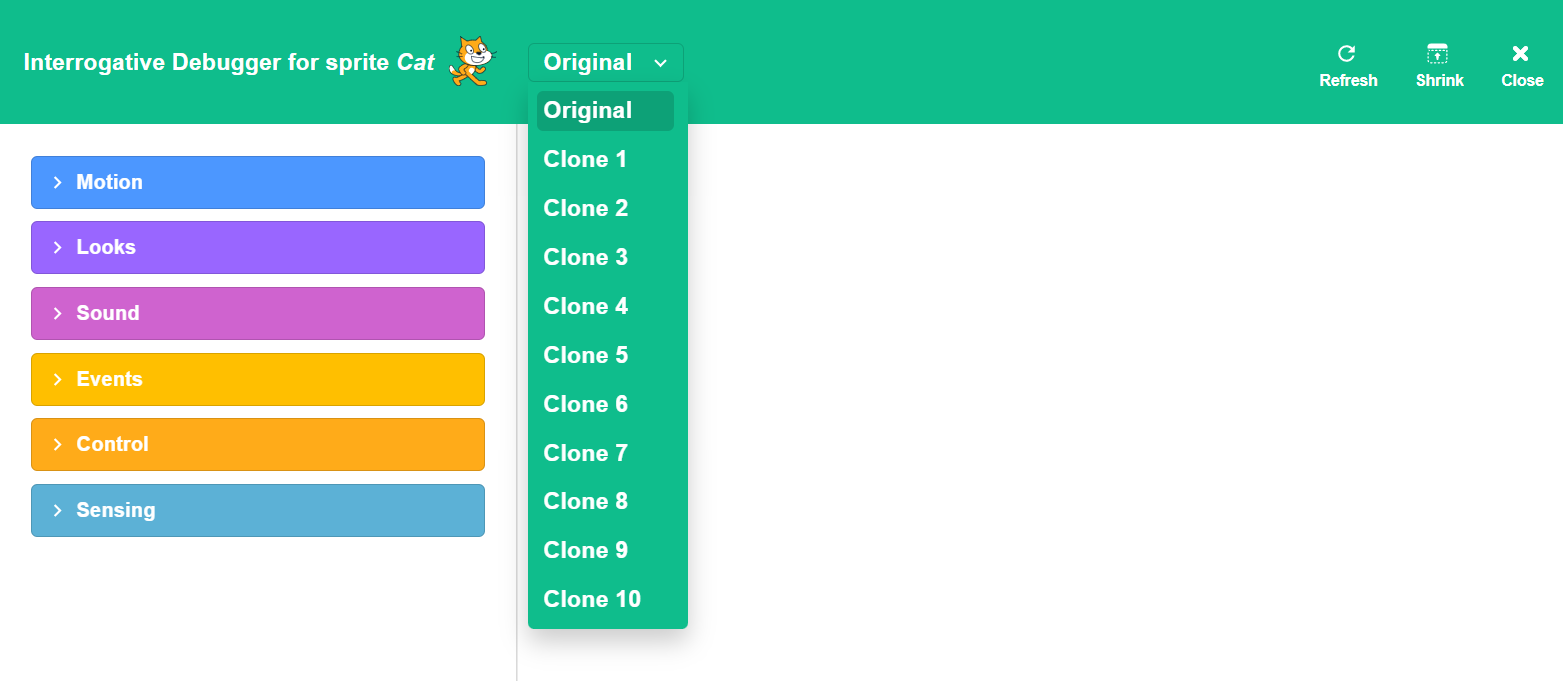}
  \vspace{-3em}
  \caption{Drop down menu to select the sprite instance of interest.}
  \vspace{-1em}
  \label{fig:clone-selection}
\end{figure}

The context menu option shown in \cref{fig:open-target-questions} can
be used to ask questions about the behavior of targets (i.e., sprites or
 the stage).
Since a sprite may have clones, users can choose which 
 instance they want to ask questions about using a drop
 down menu in the dialog's header (\cref{fig:clone-selection}).
%
An overview of all target questions can be found in
\cref{tab:target-questions}. Each of these questions relates to
either a certain type of block, or a change in attribute. For example,
if the code contains 
\setscratch{scale=0.4}\begin{scratch}
  \blockmove{point towards \ovalmove*{object}} \end{scratch} blocks,
\toolname generates the question \textmessage{``Why
  did{\color{lightgray}{n't}} <sprite> point towards
  \textmessage{<object>}?''}
for each target object (i.e., another sprite or the mouse-pointer).
The question is positive (\WhyDid) if at least one of the related blocks was
executed, and negative (\WhyDidnt) otherwise.

\begin{table}[htb]
\caption{Block questions.}
\vspace{-1em}
\myfloatalign
\begin{tabularx}{\columnwidth}{X}
\toprule
\textsc{\textcolor[HTML]{D0112A}{Execution}} \\ 
\textmessagetable{Why did{\color{lightgray}{n't}} the block execute?} \\
\textmessagetable{When did the block execute?} \\
\midrule
\textsc{\textcolor[HTML]{4C97FF}{Motion}} \\
\textmessagetable{Why did} \setscratch{scale=0.65}\ovalmove{x position} \textmessagetable{have the value <value>?} \\
\textmessagetable{Why did} \setscratch{scale=0.65}\ovalmove{y position} \textmessagetable{have the value <value>?} \\
\textmessagetable{Why did} \setscratch{scale=0.65}\ovalmove{direction} \textmessagetable{have the value <value>?} \\
\midrule
\textsc{\textcolor[HTML]{9966FF}{Looks}} \\ 
\textmessagetable{Why did} \setscratch{scale=0.65}\ovallook{size} \textmessagetable{have the value <value>?} \\
\textmessagetable{Why did} \setscratch{scale=0.55}\ovallook{costume \selectmenu{...}} \textmessagetable{have the value <value>?} \\
\textmessagetable{Why did} \setscratch{scale=0.55}\ovallook{backdrop \selectmenu{...}} \textmessagetable{have the value <value>?}  \\
\midrule
\textsc{\textcolor[HTML]{59C059}{Operators}} \\ 
\textmessagetable{Why did{\color{lightgray}{n't}} the condition} \setscratch{scale=0.6}\booloperator{\ovalnum{} < / = / > \ovalnum{}} \textmessagetable{evaluate to true?} \\
\textmessagetable{Why did{\color{lightgray}{n't}} the condition} \setscratch{scale=0.6}\booloperator{\ovalnum{} contains \ovalnum{} ? } \textmessagetable{evaluate to true?}  \\
\midrule
\textsc{\textcolor[HTML]{5CB1D6}{Sensing}} \\ 
\textmessagetable{Why did{\color{lightgray}{n't}} the condition} \setscratch{scale=0.6}\boolsensing{touching \selectmenu{object} ? } \textmessagetable{evaluate to true?} \\
\textmessagetable{Why did{\color{lightgray}{n't}} the condition} \setscratch{scale=0.6}\boolsensing{touching color \ovalnum{} ?} \textmessagetable{evaluate to true?} \\
\textmessagetable{Why did{\color{lightgray}{n't}} the condition} \setscratch{scale=0.6}\boolsensing{\ovalnum{} is touching \ovalnum{} ?} \textmessagetable{evaluate to true?} \\
\midrule
\textsc{\textcolor[HTML]{FF8C1A}{Variables}} \\ 
\textmessagetable{Why did{\color{lightgray}{n't}} <variable> have the value <value>?} \\
\midrule
\textsc{\textcolor[HTML]{FF661A}{Lists}} \\ 
\textmessagetable{Why did <list> have the value <value>?} \\
\textmessagetable{Why didn't <list> contain <item>?} \\
\bottomrule
\end{tabularx}
\label{tab:block-questions}
\vspace{-1em}
\end{table}

\subsubsection{Block questions}

\begin{figure}[htb!]
  \centering
  \includegraphics[width=0.7\columnwidth]{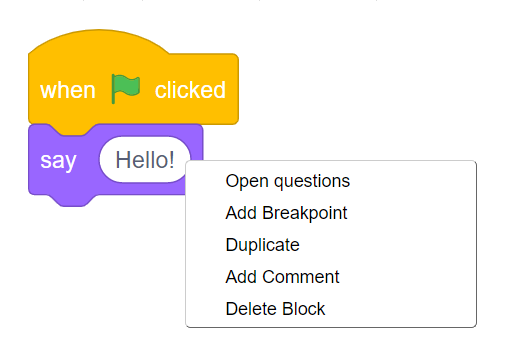}
  \vspace{-2em}
  \caption{Extended context menu of a block.}
  \label{fig:open-block-questions}
  \vspace{-1em}
\end{figure}

A context menu option \textmessage{``Open questions''} on blocks
(\cref{fig:open-block-questions}) can be used to ask about their
execution and contained reporter and Boolean blocks.
As a block may be executed multiple times,
all its executions are numbered
and selectable in a drop down menu
similar to
\cref{fig:clone-selection}.
%
Block questions are generated by iterating over the blocks of a
selected target, and instantiating all applicable questions of
those in \cref{tab:block-questions} based on whether each block
is contained in the execution trace and its value for Boolean and reporter blocks.


\subsection{Answers}
\label{sec:answers}

\begin{figure}[h]
  \centering
  \includegraphics[width=0.7\columnwidth]{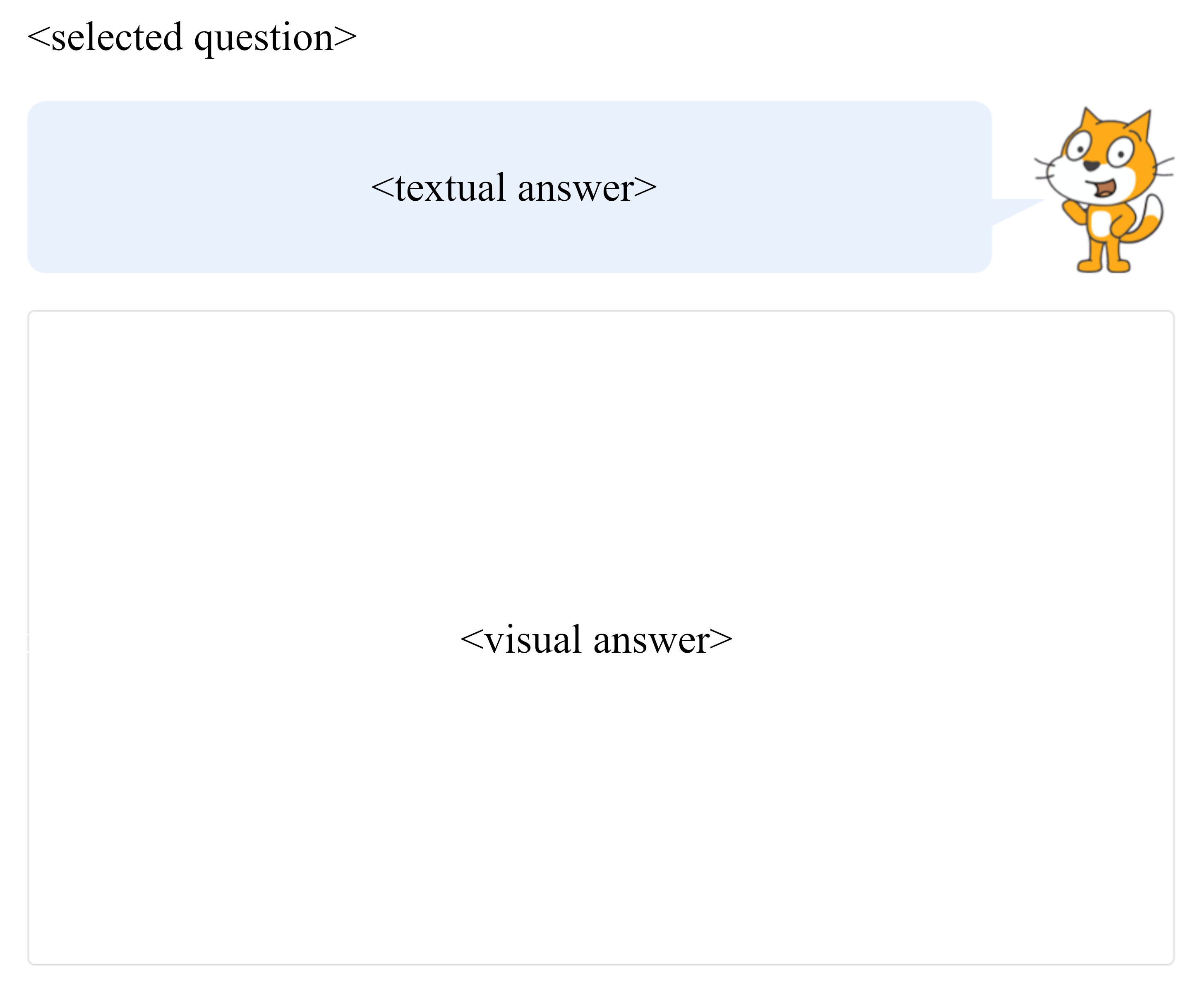}
  \vspace{-1em}
  \caption{Answer structure; examples are shown in \cref{fig:answer-position} and \cref{fig:answer-condition}.}
  \label{fig:answer-structure}
  \vspace{-1em}
\end{figure}

In contrast to \whyline~\cite{AliceWhyline}, which visualizes answers
using a graph structure representing data and control flow causality
over time, \toolname simplifies answers by separating causality and
temporality: Each answer explains the program state selected with the
slider (\cref{sec:control}) or the occurrence of a specific event or
block execution.
After selecting a question, the answer is visualized on the right side
of the dialog using the structure depicted in
 \cref{fig:answer-structure},
with a textual answer in a speech bubble
explained by a cat (the \Scratch mascot), and a visual answer 
in the form of an answer graph or visualization of the program state.


\subsubsection{Block execution answers}\label{subsec:block_anwers}

The (non-)execution of a block is based on its control dependencies,
which in \Scratch can be control blocks, hat blocks triggered by the
user, and hat blocks triggered programmatically. For an executed block
\toolname shows a graph of all control dependencies causing the
execution of the block; for unexecuted blocks the graph visualizes
every possible way the block would execute and why it did not
occur. Thus, the answer graph is a subgraph of the \ac{CDG} containing
the block of interest and some or all of its transitive predecessors.
The answer graph is created by first removing all circular
dependencies in the \acs{CDG}, since they can easily be confusing and
are not necessary to understand why a block was executed or not. Then,
starting from the target block the CDG is traversed backwards for all
transitive dependencies that do not represent user events, copying
each traversed node and edge to the answer graph.

\begin{figure}[htb]
  \centering
  \subfloat[All blocks executed.]{\includegraphics[height=0.45\columnwidth]{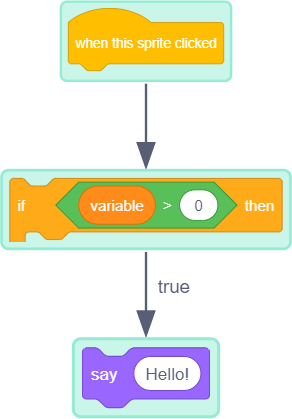}\label{fig:answer-graph-condition-1}}
  \hfill
  \subfloat[Partially executed.]{\includegraphics[height=0.45\columnwidth]{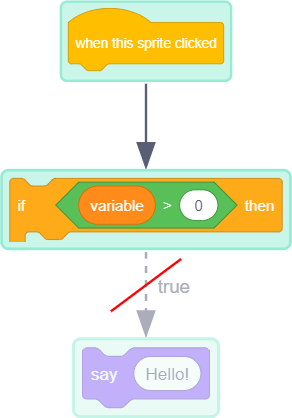}\label{fig:answer-graph-condition-2}}
  \hfill
  \subfloat[No block executed.]{\includegraphics[height=0.45\columnwidth]{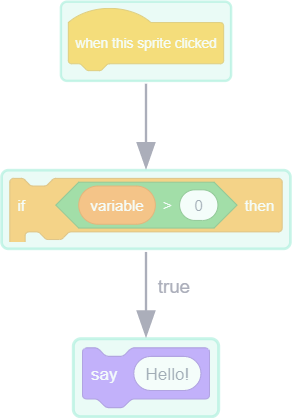}\label{fig:answer-graph-condition-3}}
  \hfill
  \caption{Different answer graphs for the say block 
  depending on whether the contained blocks were executed.}
  \label{fig:answer-graph-condition}
\end{figure}

The resulting graph is rendered using \Scratch blocks as nodes, and 
each block has a
colored margin indicating its associated target.
%
If a node represents a block that was never executed the block has a
reduced opacity of 0.5. The same holds for edges, which count as
executed if their source and target nodes were
executed. \Cref{fig:answer-graph-condition} shows the same graph with
all blocks executed, some blocks executed and none executed. If the
source of an edge is a control block containing a condition, the
condition's value required to flow this direction is used as label for
the edge. If the control block was executed, the visualization depends
on the condition's traced value: If it is not the required one, then
the edge is dashed and crossed out using a red line
(\cref{fig:answer-graph-condition-2}).
If the execution should have caused the execution of another block,
but that block was not executed, the execution was interrupted, which
is indicated with a crossed out edge labeled with
\textmessage{``execution paused / stopped''}.

\begin{figure}[htb]
  \centering
  \subfloat[Tooltip for a reporter block (\setscratch{scale=0.65}\ovalvariable{variable}).]{\includegraphics[width=0.36\columnwidth]{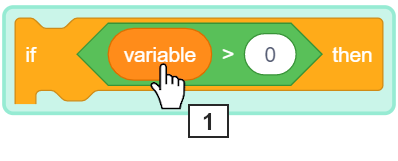}\label{fig:tooltips-oval}}
  \hspace{6mm}
  \subfloat[Tooltip for a Boolean block (\setscratch{scale=0.65}\booloperator{\ovalnum{} > \ovalnum{}}).]{\includegraphics[width=0.36\columnwidth]{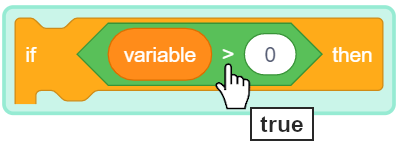}\label{fig:tooltips-bool}}
  \caption{Tooltips showing the traced value of parameter blocks.}
  \label{fig:tooltips}
\end{figure}

Hovering the mouse cursor over a reporter block
(\cref{fig:tooltips-oval}) or a Boolean block
(\cref{fig:tooltips-bool}) displays the value this block had during
execution.
Clicking a node in the answer graph opens the interrogative debugger
for that block, and if it was executed multiple times, the relevant
execution is selected.

Textual answers are generated using the answer graph. For positive
questions this answer explains every control dependency responsible
for a block's execution. Starting with the entry node of the graph, a
message is generated by adding \textmessage{<reason$_1$>} to describe
the control dependency, and then iteratively appending explanations
\textmessage{<reason$_i$>} of the node's successors with appropriate
conjunctions. For this, \toolname defines positive and negative
parametrizable text templates for each block.
For example, for the graph in \cref{fig:answer-graph-condition-1} the
answer is \textmessage{``The block was executed, because the sprite
  Cat was clicked and afterwards the condition
}\setscratch{scale=0.6}\booloperator{\ovalvariable{variable} >
  \ovalnum{0}} \textmessage{was true.''}

The textual explanation why a block was not executed is based on the
reason in the answer graph that prevented the execution from flowing
towards the block of interest, i.e., the edge where the source block
was executed and the target was not.  The answer then consists of
\textmessage{``The block wasn't executed, because <reason>.''}
If the execution to the target node flowed in the correct direction but
was interrupted, 
\textmessage{<reason>} is \textmessage{``the execution was stopped /
  paused''}. Otherwise, if the execution did not flow in the correct
direction due to a condition not having the required value,
\textmessage{<reason>} is \textmessage{``the condition <condition>
  wasn't <requiredValue>''}.
Since events can depend on multiple blocks or on user input, the
\textmessage{<reason>} for nodes depending on events is set to a
generic message specified for each type of event.
If the node that interrupted the execution from flowing towards the
target block is not its immediate predecessor, the template for the
answer is \textmessage{``The block wasn't executed, because <reason>
  and therefore all subsequent blocks that could lead to the execution
  of the block were not executed.''}, since
resolving \textmessage{<reason>} might still not guarantee the
execution of the target block.


\begin{figure}[h]
  \centering
  \includegraphics[width=0.18\columnwidth]{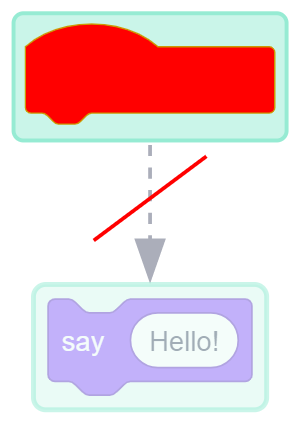}
  \vspace{-1em}
  \caption{Answer graph for an unreachable say-block.}
  \label{fig:unreachable-answer}
  \vspace{-1em}
\end{figure}

Blocks that are not executed because they are unreachable (dead code)
result in \textmessage{``The block wasn't executed because it is not
  reachable!''}, and the visualization shown in
\cref{fig:unreachable-answer}.


\subsubsection{Block execution time answers}

\begin{figure}[h]
  \centering
  \includegraphics[width=0.7\columnwidth]{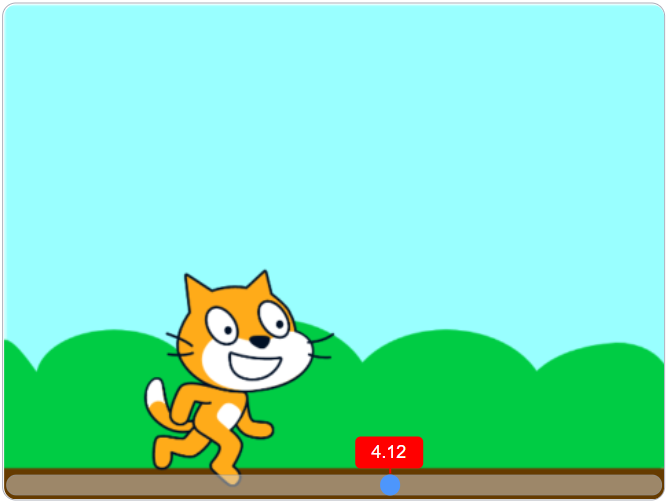}
  \vspace{-1em}
  \caption{Visual answer showing the execution time of a block.}
  \label{fig:execution-time-answer}
  \vspace{-1em}
\end{figure}

To answer the question \textmessage{``When did the block execute?''}
\toolname extracts the timestamps of the first trace entry and the
selected block execution, and calculates their difference in
seconds. If the first trace entry recorded the execution of the
\setscratch{scale=0.3}\begin{scratch} \blockinit{when \greenflag
    clicked}\end{scratch} block, the answer message is
\textmessage{``The block was executed <elapsedTime> seconds after
  clicking the green flag.''} and otherwise it is \textmessage{``The
  block was executed <elapsedTime> seconds after starting the
  recording.''} The visual part of the answer is a picture of the
stage rendering the state of the program when the selected block
execution took place (\cref{fig:execution-time-answer}).
Below the picture a slider with a fixed value indicates the time of
the block execution, with the calculated temporal difference shown as
a red label. 

\subsubsection{Target behavior answers}

Questions about target behavior refer to the 
execution of one or multiple blocks.  For example, a question like
\textmessage{``Why did{\color{lightgray}{n't}} <sprite> say
  <message>?''} can be transformed to \textmessage{``Why
  did{\color{lightgray}{n't}} a say <message> block of <sprite>
  execute?''}
%
The general procedure to answer positive questions about target
behavior is to extract all blocks that caused the behavior and then
answer why each of these was executed.
If
there are multiple graphs (e.g., for ambiguous control flow), then
they are merged.
 Starting at the end of the trace causes the graph
nodes to visualize values of the latest relevant execution.
%
%

\begin{figure}[h]
  \centering
  \includegraphics[width=0.4\columnwidth]{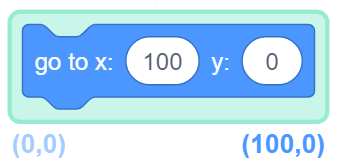}
  \vspace{-1em}
  \caption{Graph node visualizing the position of a sprite before and after the relevant execution of the node's block.}
  \label{fig:position-before-after}
\end{figure}

Depending on the question and related block, relevant attribute values
before and after the execution of the block are added to the graph's
node, for example the position of a sprite before (left) and after (right) the
block's execution
(\cref{fig:position-before-after}). The same holds for blocks changing
the attributes direction, size, costume and backdrop.

The textual answer is of the form \textmessage{``<behavior>, because
  <reason$_1$>, then <reason$_2$>, next <reason$_3$>, then
  <reason$_4$>, ..., next <reason$_x$> and therefore the <type> block
  was executed.''}, where \textmessage{<behavior>} is a text
describing the target's behavior the question asks about,
\textmessage{<type>} is the type of the block causing the behavior,
and \textmessage{<reason$_i$>} are based on the control dependencies.
If more than one block's execution caused the question's behavior,
then there are multiple graphs, in which case a question mark button
is shown next to each leaf node. 
Initially, the general answer message \textmessage{``<behavior>,
  because the execution of <count> <type> blocks caused this
  behavior. Do you need an explanation for a block? Then click on the
  ? next to it!''} is displayed. Clicking a question mark button
next to a node highlights the node
and displays its answer message. 

For negative questions 
the strategy is to find all blocks that could lead to this behavior,
and then explain why each of these did not cause it.
If a block was not executed, the answer graph is used to show why not
and to determine the answer message of the form
\textmessage{``<behavior>, because <reason> and therefore the <type>
  block was not executed.''}, where \textmessage{<behavior>} is a text
describing the queried unobserved behavior, and \textmessage{<type>}
is the type of the block that could lead to the behavior;
\textmessage{<reason>}
is created as described in \cref{subsec:block_anwers}.
Unobserved target behavior can also be explained by (1) blocks
attached to a \setscratch{scale=0.3}\begin{scratch}
  \blockinitclone{when I start as a clone}\end{scratch}, which is
never executed by an original sprite; (2) blocks attached to the block
\setscratch{scale=0.3}\begin{scratch} \blockinit{when \greenflag
    clicked}\end{scratch} but queried for a clone, which can never be
executed since clicking the green flag deletes all clones; and (3)
events that occurred outside the lifetime of a clone. All of these
cases have dedicated answers.



\subsubsection{Reporter block answers}

\begin{figure}[h]
  \centering
  \subfloat[Graph explaining the data dependencies starting at the latest block overwriting the value.]{\includegraphics[width=0.36\columnwidth]{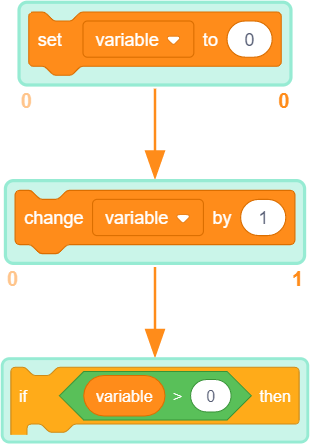}\label{fig:data-dependence-graph-1}}
  \hspace{15mm}
  \subfloat[Graph explaining the data dependencies starting at the initial value.]{\includegraphics[width=0.36\columnwidth]{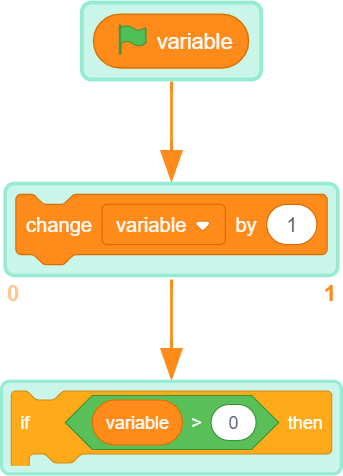}\label{fig:data-dependence-graph-2}}
  \vspace{-1em}
  \caption{Visual answers for a positive question asking about the value of a variable.}
  \label{fig:data-dependence-graph}
\end{figure}

Questions about reporter blocks are only available if the object of
interest is an executed block containing them. The answer graph
visualizes blocks causing or preventing the reporter block from having
a certain value.
%
To answer the positive question \textmessage{``Why did <variable> have
  the value <value>?''} this graph contains the data dependencies
responsible for the variable's recorded value \textmessage{<value>}.
In order to distinguish between data and control dependence edges,
data dependence edges are colored using the relevant value's color
(e.g., orange for variables); values before and after each block are
shown beneath the node (\cref{fig:data-dependence-graph-1}).
If the graph's entry node represents the initial value, this node is
visualized using
\inlineFigure{0.3}{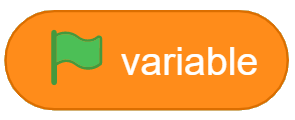} if
the first traced block is \setscratch{scale=0.3}\begin{scratch}
  \blockinit{when \greenflag clicked} \end{scratch}
(\cref{fig:data-dependence-graph-2}), or else using
\inlineFigure{0.3}{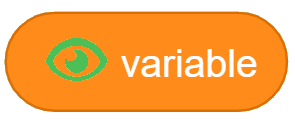}.
%
%
An answer message is generated by traversing the graph from the entry
node, appending a pre-defined message \textmessage{<reason$_i$>} for
each block's effects, and thus generating the answer
\textmessage{``<variable> had the value <value>, because <reason$_1$>,
  then <reason$_2$>, then <reason$_3$>, ... and afterwards
  <reason$_x$>.''}

The negative question \textmessage{``Why didn't <variable> have the
  value <value>?''} is generated, if the program contains at least one
block setting the variable's value to \textmessage{<value>}, but the
value traced during the selected block execution is different. First,
all relevant blocks setting the variable to \textmessage{<value>} are
determined.
If none of these was executed, then the corresponding answer graphs
are rendered side by side, together with a general message (with
detailed messages accessible with the question mark button.)
%
If at least one set block was executed and the variable had the value
at some point but was changed before the selected block execution, the
answer is \textmessage{``The variable was set to <value>, but changed
  afterwards.''}, with a graph visualizing all block executions
changing the variable's value.

\subsubsection{Boolean operator block answers}

The question \textmessage{``Why did{\color{lightgray}{n't}} the
  condition \setscratch{scale=0.6}\booloperator{\ovalnum{} <operator>
    \ovalnum{}} evaluate to true?''} is answered with the values of
all reporter blocks within the Boolean operator block. For
example, for \textmessage{``Why did the condition}
\setscratch{scale=0.6}\booloperator{\ovalvariable{A} <
  \ovalvariable{B}} \textmessage{evaluate to true?''}  the values of
\setscratch{scale=0.6}\ovalvariable{A} and \ovalvariable{B} are
extracted from the trace. Assuming \ovalvariable{A} =
0 and \ovalvariable{B} = 1, the answer is:
\textmessage{``The condition evaluated to true, because} 
\ovalvariable{A} \textmessage{had the value} \ovalnum{0}, 
\ovalvariable{B} \textmessage{had the value} \ovalnum{1} 
\textmessage{and therefore} \setscratch{scale=0.6}  \booloperator{\ovalnum{0} < \ovalnum{1}} \textmessage{is true.''}
%
If there are no reporter blocks, the answer
is \textmessage{``The condition is always true / false, because none
  of the blocks is a \ovalvariable{variable}''}.

\begin{figure}[tb]
  \centering
  \subfloat[Cat touching the butterfly.]{\includegraphics[width=0.465\columnwidth]{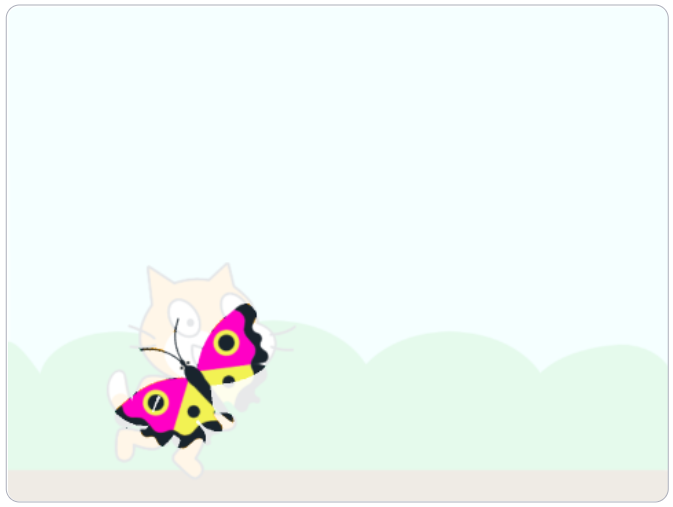}\label{fig:touching_answer_e}}
  \hfill
  \subfloat[Cat not touching the butterfly.]{\includegraphics[width=0.465\columnwidth]{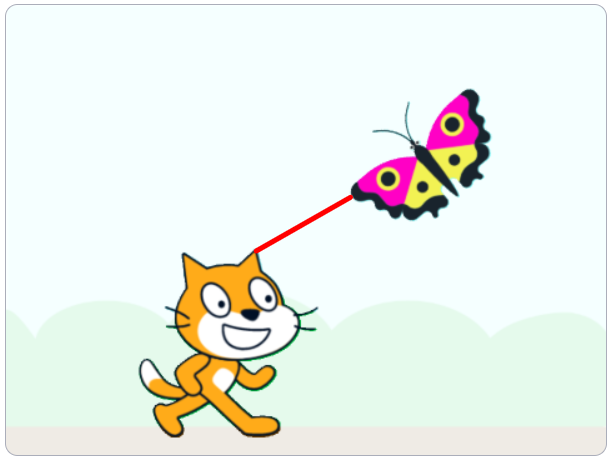}\label{fig:touching_answer_f}}
  \vspace{-1em}
  \caption{Visual answers explaining the values of a Boolean touching block evaluating to true and false.}
  \label{fig:touching_answer}
  \vspace{-1em}
\end{figure}

In order to answer \textmessage{``Why did{\color{lightgray}{n't}} the
  condition} \setscratch{scale=0.6}\boolsensing{touching ?}
\textmessage{evaluate to true?''} questions,
a picture of the stage in the relevant state is used to visualize the
reason for the condition's value. Each existing different kind of
\setscratch{scale=0.6}\boolsensing{touching ?} block checks if a set
of positions $P_{A}$ is touching another set of positions $P_{B}$.
For \setscratch{scale=0.5}\boolsensing{touching \selectmenu{object} ?}
blocks $P_{A}$ are the positions of the sprite containing the block
and $P_{B}$ are the positions of the object selected in the drop down
menu, which could be another sprite (i.e., all non-transparent pixels
within that sprite), the stage's edges (i.e., positions having an
x-value of $\pm$240 or an y-value of $\pm$180 ) or the mouse-pointer
position.
\setscratch{scale=0.6}\boolsensing{ touching color
  \pencolor{magenta} ? } blocks compare the positions $P_{A}$ of the
sprite containing the block with all positions $P_{B}$ on the stage
having the selected color.
For \setscratch{scale=0.6}\boolsensing{
  \pencolor{orange} is touching \pencolor{magenta} ? } blocks $P_{A}$
are all positions of the first selected color and $P_{B}$ are all
positions of the second one.
If the positions $P_A$ and $P_B$ are touching each other,
all overlapping positions $P_A \cap P_B$ are highlighted
with decreased opacity of all other positions
$P_{\text{Stage}} \setminus P_A \cap P_B$.
\Cref{fig:touching_answer_e} shows an example of the
\setscratch{scale=0.6}\boolsensing{ touching \selectmenu{Butterfly} ?
} evaluating to \textcode{true}; other types of
\setscratch{scale=0.6}\boolsensing{touching ?} blocks are visualized
similarly.  In addition, the answer message \textmessage{``When the
  block was executed, <A> touched <B> as shown in the picture.''} is
shown. 

For false conditions, the distance of each position in $P_A$ to each
position in $P_B$ is calculated and $p_A \in P_A$ and $p_B \in P_B$
are selected, such that the distance between $p_A$ and $p_B$ is
minimal. The distance is visualized in the picture by drawing a red
line between $p_A$ and $p_B$ and highlighting the positions $P_A$ and
$P_B$ by decreasing the opacity of all other positions
$P_{\text{Stage}} \setminus P_A \cup P_B$.
\Cref{fig:touching_answer_f} shows an example of the
\setscratch{scale=0.6}\boolsensing{ touching \selectmenu{Butterfly} ?
} evaluating to \textcode{false}.
Then, the answer message
\textmessage{``The distance from <A> to <B> was <distance> when the
  block was executed.''} is generated and displayed in the speech
bubble. It can happen that a list of positions is empty because a
sprite is invisible or a color does not exist on the stage. In this
case no distance can be calculated, and the answer given is:
\textmessage{``<sprite> could not be touched, because it was invisible
  when the block was executed!''} or \textmessage{``The color <color>
  did not occur when the block was executed! Try to select the desired
  color with the color picker.''}




\section{Evaluation}

In order to provide initial insights on the usability of
\toolname, our evaluation aims to answer the
following research questions:

\noindent\textbf{RQ1: } Do teachers consider \toolname to be useful? \\
\noindent\textbf{RQ2: } Can children find and fix faults using \toolname?

\subsection{RQ1: Usability Study with Teachers}

\subsubsection{Experimental Setup}

We surveyed 11 secondary school computer science teachers during a
teacher training at the University of \emph{blinded}.  All 11 teachers
already had prior experience with \Scratch, and the teacher training
included a \Scratch programming activity prior to the survey. For the
survey itself, we first demonstrated the debugger's functionality for
30 minutes, then let the teachers use \toolname to debug nine faulty
programs for one hour, and finally asked them to complete a survey
based on the \textit{\ac{TAM}}, a common instrument to predict the
acceptance and usage of a technology~\cite{tam}. The model specifies
26 survey items of which we used 22, because the others were not
relevant or applicable. All items were measured on a 7-point Likert
scale ranging from strongly disagree to agree. At the end, the survey
contained an optional open question for textual feedback. The full
survey and the responses are available in the replication
package\footnote{\url{https://figshare.com/s/68281791e01be5179699}}.

\subsubsection{Threats to validity}

The survey is small and subject to threats to \emph{external validity}
concerning the generalization of results.  Threats to \emph{internal
  validity} arise as we introduced \toolname within 30 minutes and
afterwards the teachers used it for one hour only, which may not be
sufficient to identify all strengths or weaknesses. While the teachers
debugged programs using \toolname, they did not experience it while
\emph{teaching} debugging, thus implying a threat to \emph{construct
  validity}. Nevertheless, all participants were teachers in practice
with experience in teaching, and the results give a good first
impression on the usability and usefulness of \toolname.

\subsubsection{Results}

\begin{figure}[tb]
  \centering
  \includegraphics[trim={0 0 1.6cm 0},clip,width=\columnwidth]{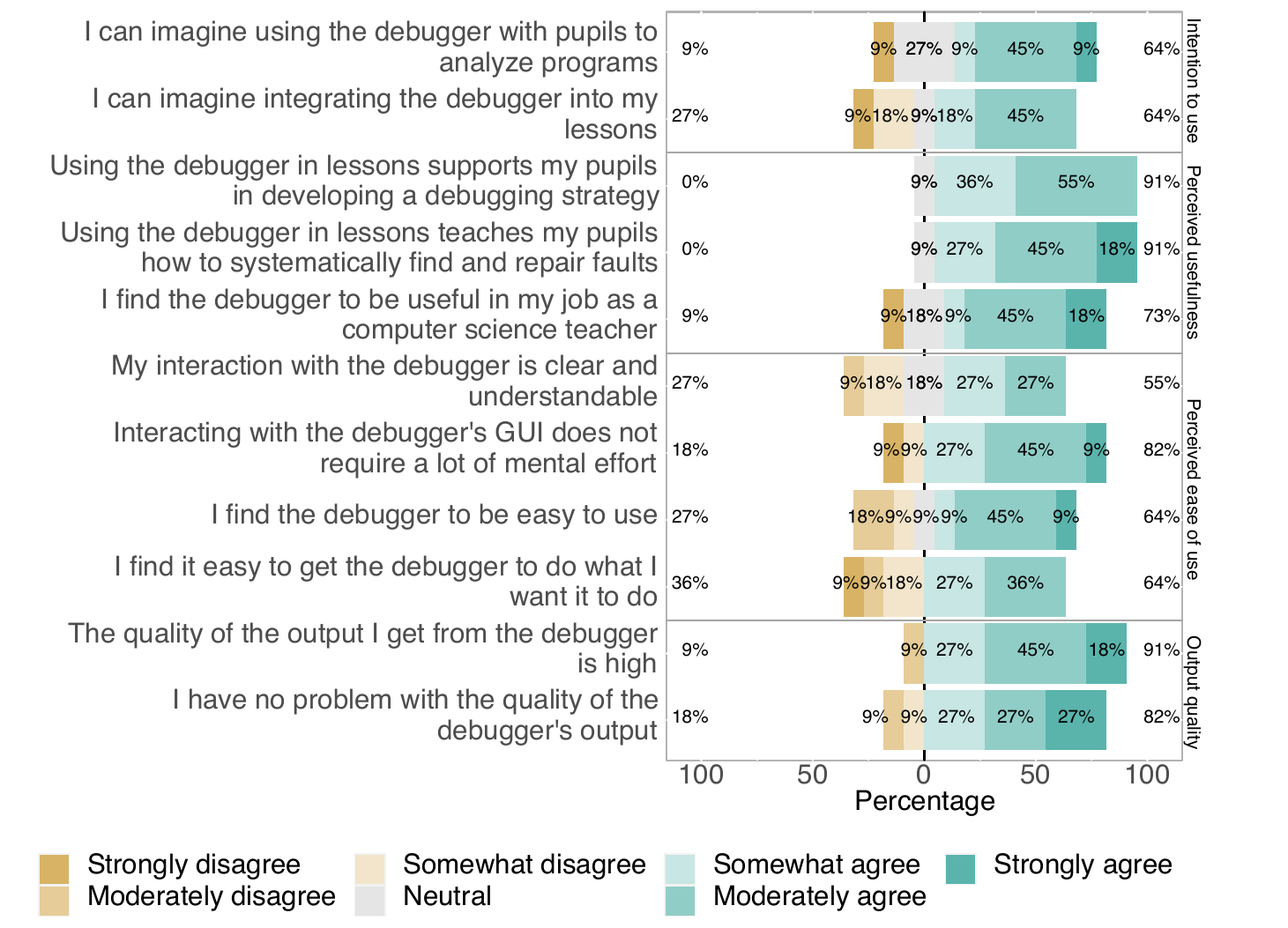}
  \vspace{-2em}
  \caption{\label{fig:survey} Teacher responses to \textit{\ac{TAM}}
    questions on intention to use, perceived usefulness, perceived
    ease of use, and output quality. Percentages aggregate all levels
    of agreement or disagreement.}
\end{figure}

\textit{\ac{TAM}} contains questions in different categories; for
space reasons, \cref{fig:survey} only shows the questions of the
categories relevant for RQ1 (intention to use, perceived usefulness,
perceived ease of use, and output quality).
The responses indicate that the majority of teachers considers
\toolname to be useful, especially in supporting pupils in developing
a debugging strategy (91\% agree somewhat or moderately). The output
of \toolname is understandable for almost all teachers (91\% agree the
quality of the output is high, and 82\% had no problems with the
output), and more than half (64\%) can imagine integrating it into
their lessons. Six out of the 11 teachers provided free-text feedback,
which was positive in five cases, for example, ``\textit{The debugger
  is a very nice tool}'' or ``\textit{That’s a nice idea}''. They also
commented that the experience of teaching using the debugger would be
helpful to provide better feedback.

However, the study also revealed drawbacks of \toolname. Questions
regarding ease of use are rated comparatively low (\cref{fig:survey}),
and four of the teachers commented on the user interface in their
free-text answers, such as
``\textit{Sometimes it was difficult to understand where the block
  shown in the answer graph is located in the program.}'' and
``\textit{The questions are slightly hidden, because they are only
  accessible via the context menu.}''  We plan to improve usability,
for example by adding direct navigation from answers to corresponding
scripts, or a questionmark-cursor to click directly on
objects, even on the rendered stage.
One teacher rated almost all items negatively, and justified this with
a general aversion of \Scratch independently of the debugger. Another
teacher praised the debugger in the free-text response but at the same
time explained low ratings with being unsure whether to use \Scratch
in lessons at all. More generally, \cref{fig:survey} shows that even
though the perceived usefulness is high, the intention to use is
visibly lower, which confirms issues in how debugging is perceived in
education~\cite{michaeli2019current}.

\summary{RQ 1}{Teachers believe that \toolname is useful and
  understandable, but suggest improving the user
  interface.}

\subsection{RQ2: Usability Study with Children}





\subsubsection{Study participants} We conducted the study at a secondary
school with six school classes (two year 6 aged 12, one
 year 7 aged 13, two year 9 aged 15, and one year 10 aged 16)
with a total of $125$ pupils. Prior to our study, the computer science
teacher of these classes implemented the same \Scratch introduction in
regular lessons for all classes, in which all programming basics
required for our study (e.g., loops, branches, message passing, sprite
cloning) were first explained, and then practiced by the pupils using
a detailed task of implementing a game covering all these concepts.
%
%

\begin{table}[t]
  \caption{\label{tab:tasks}Debugging task statistics}
  \vspace{-1em}
  \begin{minipage}[t]{\columnwidth}
    \scriptsize Difficulty counts programming concepts (easy = 2;
    medium = 3; hard = $>$ 3); B = Blocks; S = Scripts; ICC =
    inter-procedural cyclomatic complexity; Bugs refer to common
    \Scratch bug patterns~\cite{fradrich2020common}.
    \end{minipage}
  \resizebox{\columnwidth}{!}{%
      \begin{tabular}{ll@{}rrrrl@{}r}
    \toprule
    Task & Difficulty & B & S & ICC & Tests & Bug & Data points\\
    \midrule
    1 & easy   &  23 & 4  & 7  & 3 & Type Error & 125 \\ 
    2 & easy   &  27 & 4  & 9  & 3 & Missing loop sensing & 125 \\ 
    3 & med. &  65 & 11 & 26 & 6 & Message never sent & 123 \\ 
    4 & med. &  37 &  4 & 11 & 3 & Position equals check & 122 \\ 
    5 & med. &  35 &  6 & 13 & 1 & Forever inside loop & 110 \\ 
    6 & med. &  48 &  9 & 20 & 6 & Message never received & 103 \\ 
    7 & hard   & 104 &  9 & 35 & 2 & Interrupted loop sensing & 95 \\ 
    8 & hard   &  48 &  6 & 16 & 3 & Missing clone call & 79 \\ 
        \bottomrule
        \end{tabular}%
  }
\end{table}

\subsubsection{Experiment tasks}
A debugging task in our study consists of a faulty \Scratch project
with a textual description of (1) expected behavior of the program and
(2) the incorrect behavior, and (3) their task, which is to ``fix the
faulty behavior''. \Cref{tab:tasks} shows details of the individual
tasks in terms of the difficulty estimated using the number of
programming concepts (e.g., loops, conditions, message passing, sprite
cloning, ...) included in a program (easy = 2, medium = 3, hard $>$
3), the number of blocks and scripts, and the inter-procedural
cyclomatic complexity. Each task was created to contain exactly one
bug, and the bugs are based on common bug patterns in
\Scratch~\cite{fradrich2020common}. In addition we created four
introductory debugging tasks, two of which introduce bugs into the
program created by the pupils in their \Scratch preparation lesson,
and two easy bugs similar to the main experiment tasks.
To increase the number of data points, for each task there are two
versions with identical code and bug, but slightly changed theme and
storyline.

\subsubsection{Experiment procedure}

The study itself covered two two-hour lessons, each consisting of an
(1) introduction with explanations, followed by (2) two introductory
tasks, and then (3) 40 minutes dedicated to four debugging tasks.
In the first week we taught the basics of debugging without a debugger
using teaching material adapted from existing didactic approaches for
systematic debugging with text-based
languages~\cite{michaeli2019improving}. The introductory tasks include
diagrams explaining the steps involved in systematic debugging.
After the introduction the pupils practiced debugging
using four debugging tasks.
In the second week we first demonstrated systematic debugging using
\toolname, and pupils then practiced with two introductory tasks,
where the task description included modified diagrams of the steps
involved in systematic debugging with a
debugger~\cite{michaeli2019improving}.
Afterwards, they had to solve four debugging tasks within 40
minutes. Their instructions were to fix the bugs, but they were free
to choose whether to use \toolname. We used a modified \Scratch
version~\cite{caspari2023scratchlog} with a button for indicating when
a task is completed, collecting the corresponding \textsf{.sb3} file
and time, and resulting in the data points per task shown in
\cref{tab:tasks}.

Each session consisted of one easy, two medium, and one hard task. To
collect data for all eight tasks in the second session, we used tasks
1, 3, 4, 7 for half the classes in the first session and tasks 2, 5, 6,
8 in the second session, and vice versa for the other half. To avoid
that information about tasks was exchanged between groups, we used the
alternative task versions depending on the order.

\subsubsection{Experiment analysis}
The analysis is based on the eight debugging tasks used during the
second session. For each pupil and task a modification in the \Scratch
UI logged whether they used \toolname, and we compare those who used
\toolname with those who did not.
%
To determine whether a bug was correctly fixed we created automated
tests for each task using the \Whisker~\cite{whisker} framework, which
allows specifying UI tests for \Scratch programs using JavaScript
notation. The number of tests per task is shown in \cref{tab:tasks};
note that the tests not only check if the bug was fixed, but also
check that all other behavior of the program has not been broken. We
executed the tests on each pupil solution, and count a bug as
successfully fixed if all tests pass.
We compare the fix ratios between pupils using and not using \toolname
with a Fisher exact test at $\alpha = 0.05$ and the odds ratio as
effect size measure.
The modified version of \Scratch~\cite{caspari2023scratchlog} also
keeps track of the time spent per completed task. We compare the times
between pupils using and not using \toolname with a Mann-Whitney U
test at $\alpha = 0.05$.

\subsubsection{Threats to Validity}

Threats to \emph{internal validity} result as we did not force pupils
to use a debugger, but compare those who chose to use it with those
who did not, which may be those who immediately spotted bugs or are
overwhelmed by the debugger. To reduce the influence of the latter
cause, we monitored the pupils during the study and provided feedback
to pupils who were lost.
Results may differ depending on how debugging is taught, therefore the
tasks and experiment sessions were co-designed and conducted by a
teacher. We also conducted a pilot study with another school class and
refined the debugger and teaching material based on the experience.
Results may further differ depending on programming knowledge, but all
pupils received the same introduction to Scratch and were taught by
the same computer science teacher.
Threats to \emph{construct validity} arise from measuring whether
programs were fixed and how quickly; the debugger may have helped
locating but not fixing the fault. Furthermore, in an educational
setting the learning outcome may be more important than the raw fix
ratio or performance.
Threats to \emph{external validity} arise from our sample of
pupils and faulty programs, and results may not generalize.

\subsubsection{Results}

\begin{figure}[tb]
  \centering
  \includegraphics[width=\columnwidth]{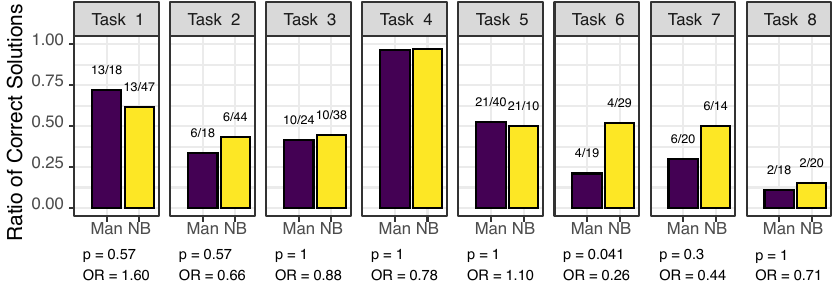}
  \vspace{-2em}
  \caption{\label{fig:ratio}Ratio of correctly fixed programs with manual debugging (Man) and with \toolname (NB).}
  \vspace{-0.8em}
\end{figure}

\begin{figure}[tb]
  \centering
  \includegraphics[width=\columnwidth]{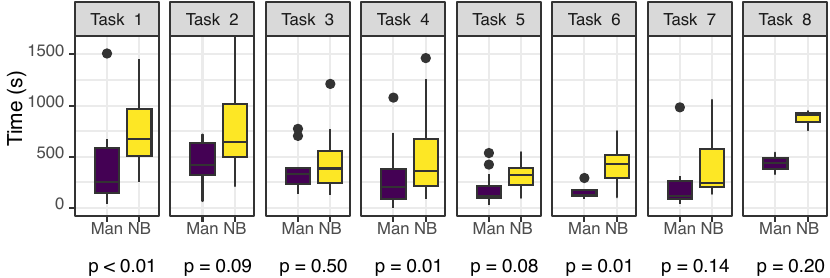}
  \vspace{-2em}
  \caption{\label{fig:time}Time spent for correct fixes.}
  \vspace{-0.8em}
\end{figure}

Overall, 82 debugging tasks were addressed using \toolname, and 65
without.
\Cref{fig:ratio} shows the ratio of pupils who correctly fixed the
bug. For six out of the eight
tasks, the ratio is higher when using \toolname, and for Task~6 the
improvement is statistically significant (Fisher exact test at
$\alpha = 0.05$). This bug consists of incorrectly nested
forever-loops, which can be detected with stepping, or by asking why
blocks or behavior outside the loops are not reached. A large
improvement can also be seen for Task 7, which contains incorrect
handling of clone generation, which can easily be understood by asking
the right question. Both of these tasks represent more complex
programs; the task where the debugger helped least (Task 1) is a very
simple program, where sprite collision is incorrectly checked only
once rather than continuously. Understanding this bug requires
understanding timing visualizations, for which deriving a fix may be
more challenging as no code is shown---which may be challenging for
children, who have been shown to focus on code rather than deducing a
causal model~\cite{kim2018debugging}.

\Cref{fig:time} shows the time spent for the cases where the bug was
successfully fixed. For all tasks the time is higher with \toolname
(significant for tasks 1, 4, and 6). On the one hand, this may be
influenced by pupils ignoring the debugger if they spot a bug
immediately. On the other hand, \toolname provides substantially more
information to process compared to the few blocks of most programs.
Interestingly, the \whyline interrogative debugger was reported to
reduce time~\cite{AliceWhyline}. However, we did not study students in
higher education, but children, which leads to unique challenges, such
as their focus on solutions themselves rather than how the program
works~\cite{kim2018debugging,vessey}. Furthermore, in a teaching
context it might actually be more productive when pupils spend more
time with debugging as this may consolidate knowledge.


\summary{RQ 2}{The effectiveness of pupils at fixing faults
  increases when using \toolname, but those who manage to fix the
  fault without the debugger tend to be quicker.}



\section{Conclusions}

Programmers who are good at debugging are more likely to be good at
programming~\cite{chmiel}, but being able to program does not
immediately result in being able to debug~\cite{kessler}.
Novice programmers in particular lack some of the necessary
skills~\cite{ducasse} for debugging, are often discouraged by errors
in their code~\cite{papert}, and struggle with
debugging~\cite{fitzgerald,murphy,Jadud2004AFL,beth,vessey}.
Although the importance of debugging for novices has been known for
decades~\cite{Soloway,papert}, most programming courses still do not
cover it in detail~\cite{carter} and debugging tools are lacking, in
particular for \Scratch. This inhibits novices, debugging education,
and educators trying to support students.

To remedy this situation we introduced \toolname, a debugger for
\Scratch providing classic, omniscient, and interrogative debugging
functionality. Our initial experiments showed promising results, but
also suggest future research on improving the techniques on which
\toolname is based, for example by narrowing down or abstracting
questions, by integrating further automated debugging techniques such
as spectrum-based fault localization~\cite{wong2016survey}, by
improving answers with fix
suggestions~\cite{fein2022catnip,price2017isnap,price2016generating,mcbroom2021survey,zhang2022repairing},
by integrating generated code explanations~\cite{sarsa2022automatic},
or by automatically generating
breakpoints~\cite{zhang2010bpgen}. There is also further need for
research on education of systematic
debugging~\cite{michaeli2019improving} since debugging is generally
not taught sufficiently, and in particular not using debugging
tools~\cite{michaeli2019current}.
%
%
%
To support education and research on debugging, all code, tasks, and
data are available at:
\mbox{\url{https://figshare.com/s/68281791e01be5179699}}

The source code of \toolname is also available here:
\begin{center}
  \mbox{\url{https://github.com/se2p/nuzzlebug}}
\end{center}

%



\section{Acknowledgments}

We thank Stephan Gram\"{u}ller and Edi Wasmeier for supporting the
studies; Phil Werli, Sebastian Schweikl, and Patric Feldmeier for
supporting the implementation; and Luisa Greifenstein and Ute Heuer
for supporting the design of questions and answers.
This work is supported by the DFG under grant FR 2955/3-1
``TENDER-BLOCK: Testing, Debugging, and Repairing Blocks-based
Programs''.

\balance

\bibliographystyle{ACM-Reference-Format}
\bibliography{related}

\end{document}